# Enhancing ReaxFF for Molecular Dynamics Simulations of Lithium-Ion Batteries: An interactive reparameterization protocol

Paolo De Angelis[1*], Roberta Cappabianca[1], Matteo Fasano[1], Pietro Asinari[1,2*], Eliodoro Chiavazzo[1*]

[1]Department of Energy "Galileo Ferraris", Politecnico di Torino, Corso Duca degli Abruzzi 24, Torino, 10129, Italy.
[2]Istituto Nazionale di Ricerca Metrologica, Strada delle Cacce 91, Torino, 10135, Italy.

*Corresponding author(s). E-mail(s): paolo.deangelis@polito.it; pietro.asinari@polito.it; eliodoro.chiavazzo@polito.it;

**Abstract**

Lithium-ion batteries (LIBs) have become an essential technology for the green economy transition, as they are widely used in portable electronics, electric vehicles, and renewable energy systems. The solid-electrolyte interphase (SEI) is a key component for the correct operation, performance, and safety of LIBs. The SEI arises from the initial thermal metastability of the anode-electrolyte interface, and the resulting electrolyte reduction products stabilize the interface by forming an electrochemical buffer window. This article aims to make a first – but important – step towards enhancing the parametrization of a widely-used reactive force field (ReaxFF) for accurate molecular dynamics (MD) simulations of SEI components in LIBs. To this end, we focus on Lithium Fluoride (LiF), an inorganic salt of great interest due to its beneficial properties in the passivation layer. The protocol relies heavily on various Python libraries designed to work with atomistic simulations allowing robust automation of all the reparameterization steps. The proposed set of configurations, and the resulting dataset, allow the new ReaxFF to recover the solid nature of the inorganic salt and improve the mass transport properties prediction from MD simulation. The optimized ReaxFF surpasses the previously available force field by accurately adjusting the diffusivity of lithium in the solid lattice, resulting in a two-order-of-magnitude improvement in its prediction at room temperature. However, our comprehensive investigation of the simulation shows the strong sensitivity of the ReaxFF to the training set, making its ability to interpolate the potential energy surface challenging. Consequently, the current formulation of ReaxFF can be effectively employed to model specific and well-defined phenomena by utilizing the proposed interactive reparameterization protocol to construct the dataset. Overall, this work represents a significant initial step towards refining ReaxFF for precise reactive MD simulations, shedding light on the challenges and limitations of ReaxFF force field parametrization. The demonstrated limitations emphasize the potential for developing more versatile and advanced force fields to upscale *ab initio* simulation through our interactive reparameterization protocol, enabling more accurate and comprehensive MD simulations in the future.

**Keywords:** Reactive Force Field, Molecular dynamics, Lithium-ion batteries, Solid electrolyte interphase

# Introduction

Lithium-ion batteries (LIBs) are and will continue to play an important role in the energy transition for energy storage and the fossil fuel replacement [1–4]. In the last decades, the growing popularity of electric vehicles, and energy storage systems [5] has fuelled the demand for batteries with greater capacity, more consistent performance over time, and improved safety [6]. While the materials used as anodes and cathodes largely determine battery capacity, the degradation phenomena that occur within the device govern battery stability and safety. One of the most important and yet poorly understood of these phenomena is the formation of a thin passivization film at the electrode-electrolyte interface, known as the solid electrolyte interphase (SEI) [7, 8]. This layer is formed by the irreversible reaction of lithium ions with the electrolyte due to the initial thermodynamic instability between the anode and electrolyte [9]. The ions used to create the SEI are subtracted from the battery's capacity. Indeed the increase in the SEI over time is one of the causes of capacity fade observed during battery charging cycles [10]. Additionally, the uncontrolled formation and growth of the SEI layer can limit the performance and life of current LIBs, and its degradation can lead to severe battery damage and uncontrolled exothermic reactions such as the battery thermal runaway [11]. Therefore, it is essential to model the SEI to understand and predict the behaviour of lithium-ion batteries and improve their performance and safety [12]. Despite the importance of SEI, it remains a conundrum due to its high reactivity and multiscale nature, which make challenging its study *in operando* and *in silico*, respectively [3, 13, 14]. The theoretical understanding of the SEI may allow control of its final characteristics and it can enable the engineering of such layer, thus possibly leading to enhanced properties including better electronic insulation, prevention or alleviation of graphite anode exfoliation [15] and dendrite formation in lithium metal batteries [16]. In addition, the modelling and subsequent theoretical understanding of SEI can contribute to significantly accelerate the discovery of new materials and electrolytes for future lithium-ion batteries [3, 14].

In the past two decades, various atomistic techniques have been utilized to study the kinetics of electrolyte dissociation in lithium-ion batteries [12]. The density functional theory (DFT) method has proven particularly useful in predicting the effects of lithium salt and additives on the final components of the solid electrolyte interphase [17–22]. Additionally, DFT has been used to investigate how the surface of the anode affects the electrolyte dissociation [23–26] and to calculate various properties of SEI products, including ionic conductivity [27–29], electron-transfer properties [26, 30, 31], and elastic modulus [32–34]. Although such simulations produce highly accurate results, they are computationally expensive. The cost of DFT simulations increases with the number of atoms in the system, typically scaling with $O(N^2)$ where $N$ is the number of atoms [35], limiting the size of the system that can be simulated to a few thousand atoms and the time interval to a few picoseconds. To observe the evolution of the SEI over extended periods, one solution is to use approximate functions $V(\mathbf{r}_i)$, known as force fields (FF), to calculate the system's energy. Indeed, the physical intuition-based functions employed in many force fields enable the replacement of the expensive Kohn-Sham functional $E[n]$ [36], based on electronic density $n$, with an approximate functional $E[V(\mathbf{r}_i)]$ that depends on the atomic configuration of the system $\mathbf{r}_i$ neglecting the degree of freedom of the electrons and thus being computationally less demanding [36]. Given the reactive nature of lithium-ion batteries, traditional FF methods, which rely on harmonic laws to approximate bonded interactions, are inadequate for their study. To overcome this limitation, the ReaxFF method [37], which utilizes the bond-order (BO) and electron equilibration method (EEM) [38] to describe the connectivity and charges of atoms, respectively, has been suggested and utilized. This enabled the use of molecular dynamics (MD) simulations to observe the electrolyte decomposition and the formation of SEI components in LIBs. For instance, Kim et al. [39] used ReaxFF optimized for C-H-O-Li species by Han at al. [40] to study the formation of the SEI on lithium metal surfaces. They observed the production and layering of organic and inorganic components in the SEI on lithium metal surface, resulting from the decomposition of an EC and Dimethyl carbonate (DMC) electrolyte mixture. Using the same force field, Guk et al. [41] demonstrated how the layer of SEI that forms on the surface of graphite prevents the percolation of the electrolyte and the anode exfoliation. After this initial work, Yun et al. [42] expanded the FF to include C-H-O-Li-Si-Li-F atoms, enabling the study of SEI formation in high-capacity batteries where graphite is replaced with silicon, which has a higher theoretical capacity ($372\,\mathrm{mA\,h\,g^{-1}}$ for $LiC_6$ and $4212\,\mathrm{mA\,h\,g^{-1}}$ for $Li_{4.4}Si$) [43]. This parameterization was then further enhanced by Wang et al. [44] Current available ReaxFFs have demonstrated success in reproducing dissociation energies and reaction kinetics. However, as discussed below in this work, they fall short in accurately describing the solid phase of the SEI components.

In this study, we propose a protocol for reparameterizing the Yun et al. ReaxFF for C-H-O-Li-Si-Li-F atoms. Our approach focuses on correcting parameters related to Li and F atoms to better capture the properties of lithium fluoride (LiF), which is one of the possible inorganic salts resulting from the formation of SEI. LiF stems mainly from fluoroethylene carbonate (FEC) decomposition: LiF-rich SEI is known to exhibit improved cycling stability in batteries [45]. Furthermore, LiF has exceptional properties such as high ion transport, electronic insulation, and mechanical properties [46], making it a crucial element in the engineering of the solid electrolyte interface, as demonstrated by Tan et al. [47]

The advancements in powerful python libraries for the management of atomistic systems, as outlined in the Table 1, have enabled the automation and orchestration of atomistic simulations using Jupyter notebooks and libraries such as ASE [48, 49], PyMatgen [50, 51], and ParAMS [52, 53]. These tools have facilitated the parametrization process and made it more efficient. By sharing the Jupyter notebooks [54, 55] and the database used in this parametrization work, Authors hope to address a current gap in the field of ReaxFF parameterization, i.e., the lack of access to raw data and datasets, which can limit the reproducibility and validity of the results presented.

This article is structured as follows: firstly, we present the reparameterization process and results of the new ReaxFF. Afterwards, we demonstrate how the enhanced force field improves the representation of the crystal structure of the LiF inorganic salt. We then proceed to test the ability of the enhanced force field to accurately describe the lithium mobility in LiF and highlight how it provides a more realistic diffusivity value for the Li atoms. Then we present a deep investigation of the results from the new ReaxFF, discussing the challenging aspects of this kind of potential. In the subsequent section, conclusion are drawn and a possible future outlook provided. Finally, the methods used are outlined in the final section.

Table 1: List of the most commonly used and useful Python tools and libraries in computational materials science. These libraries can be used in one or more phases of material *in silico* study: PreP = pre-processing, Run = simulation running, and PostP = post-processing.

|  | Description | PreP | Run | PostP |
| --- | --- | --- | --- | --- |
| ASE [48, 49] | Atomic Simulation Environment (ASE) is a Python library that provides a versatile framework for configuring, running, visualizing, and analyse atomistic simulations. Thanks to the object `Calculator` ASE provides a powerful interface to different codes. | ✓ | ✓ | ✓ |
| Pymatgen [50, 51] | Python Materials Genomics (Pymatgen) is a library for materials analysis, with a particular focus on solid-state studies and extensively used to produce and collect the data for the Materials Project (MP) [56] database. It has functions for reading and manipulating structural, thermodynamic, and electronic properties of materials. It can also handle inputs and outputs from various DFT codes and easily access the MP crystallography database via its integrated API. | ✓ |  | ✓ |

**Table 1**: List of the most commonly used and useful Python tools and libraries in computational materials science. These libraries can be used in one or more phases of material *in silico* study: PreP = pre-processing, Run = simulation running, and PostP = post-processing. (Continued)

| | Description | PreP | Run | PostP |
|---|---|---|---|---|
| AiiDA [57, 58] | Automated interactive infrastructure and Database (AiiDA) is an open-source, Python-based workflow management platform. Its main objective is to assist researchers in organizing, automating, managing, sharing, and tracking their simulations, thus enabling the reproducibility of complex workflows in computational materials science. Its plugging interface allows for the management of simulations performed with various codes, and it is designed to be used in conjunction with the Jupyter web interface, making the whole study interactive and easy to share. | ✓ | ✓ | ✓ |
| MDAnalysis [59, 60] | It is a versatile Python library mainly focused on allowing the manipulation and analysis of MD trajectories. With support for various trajectory and system configuration formats, it simplifies the process of translating data into n-dimensional NumPy [61] arrays for further analysis. | | | ✓ |
| Crystal Toolkit [62] | It is a library and web application framework for handling, manipulating, and analysing crystal structures. It is widely used in the MP database as an interactively visualizing web tool. | | | ✓ |
| Quippy [63, 64] | It is the Python high-level interface to QUantum mechanics and Interatomic Potentials (QUIP) [65]. QUIP is a set of software tools for performing MD simulations that can be used as a plugin for other programs such as LAMMPS, CP2K [66], and ASE. | ✓ | ✓ | |
| PyLammps [67, 68] | Large-scale Atomic/Molecular Massively Parallel Simulator (LAMMPS) is a powerful classical molecular dynamics simulation code. The python package, which can be installed after code compilation, allows to manage the LAMMPS simulations through the `lammps` module, a wrapper for the code's C-library API. | | ✓ | |

**Table 1**: List of the most commonly used and useful Python tools and libraries in computational materials science. These libraries can be used in one or more phases of material *in silico* study: PreP = pre-processing, Run = simulation running, and PostP = post-processing. (Continued)

|  | Description | PreP | Run | PostP |
| --- | --- | --- | --- | --- |
| PLAMS [69, 70] | The Python Library for Automating Molecular Simulation (PLAMS) is the Python interface for the commercial code Amsterdam Modeling Suite (AMS) [71]. Depending on the code used, it allows the automation of a wide range of simulations, including geometry optimization, vibrational spectroscopy, molecular dynamics, monte carlo (MC) simulations, and many other types of computational studies. |  | ✓ | ✓ |
| ParAMS [52, 53] | Parameter optimization for Atomistic and Molecular Simulations (ParAMS) is a specialized Python library included in the AMS. Its main purpose is to facilitate the optimization workflow to search parameters for empirical energy functionals such as ReaxFF and density functional based tight binding (DFTB). | ✓ |  |  |

# Results and Discussion

### ReaxFF reparameterization

In the currently available ReaxFF force fields, all the energy contributions, including the Li-F interactions, were parameterized using a database of *ab initio* simulations specifically designed to capture the dissociation energy for potential reactions in the system. As a result, both the force field of Yun et al. [42] and that of Wang et al. [44] can predict the reaction products in Si-Li batteries with acceptable accuracy, but the lack of data on the crystal properties of the inorganic salt limits the description of LiF aggregation and solid-phase transitions. To incorporate this information into the ReaxFF, a new database was created following the procedure proposed by LaBrosse et al. [73] specifically designed to build a force field for solid materials, in their case, pure cobalt crystal. The new database [74] was created using stable and metastable crystalline unit cells from the Material Project database [56] (Figure 1). Several initial configurations were created using the Pymatgen and ASE libraries. Over 300 DFT simulations were performed using those systems, and all relevant quantities, such as energies, forces, and partial charges, were extracted for the objective function. In accordance with FAIR (Findable, Accessible, Interoperable, and Reusable) principles [75], we stored all system and simulation data in an SQLite3 database using the ASE library. This database, along with metadata and clear usage instruction, has been shared on Zenodo (https://doi.org/10.5281/zenodo.7959121) and Github (https://github.com/paolodeangelis/Enhancing_ReaxFF_DFT_database) for easy accessibility and usability.

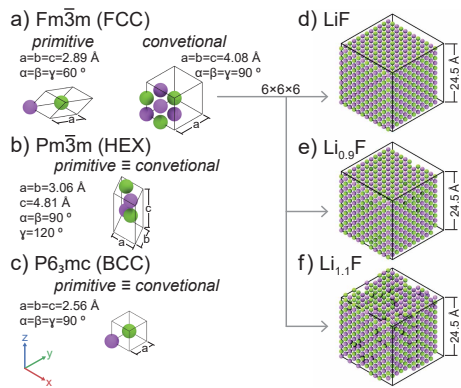

**Fig. 1** Isometric representation of unit cells used for ReaxFF reparameterization database generation and validation simulation boxes. The left column shows unit cells of LiF stable (a) and metastable phases (b, c) corresponding to face-centered cubic (FCC), hexagonal (HEX), and body-centered cubic (BCC) lattice types with space groups Fm$\bar{3}$m, Pm$\bar{3}$m, and P6$_3$mc respectively [56]. The right column displays atomistic systems used for testing the improved force field built as supercells of the stable unit cell (a): equal Li-F atoms (d), 10% vacancies at Li sites (e), and 10% interstitial Li atoms (f). The show renderings follow the Corey-Pauling-Koltun (CPK) colouring scheme [72].

The ReaxFF coefficients were optimized using the Covariance Matrix Adaptation Evolution Strategy (CMA-ES) [76] genetic algorithm. Generally speaking, this evolution algorithm searches for the optimal solution by sampling the population from a multivariate normal distribution (MVN). The generated samples are ranked based using the chosen loss function, and the best points, representing lower values for the loss function, are utilized to update the covariance and mean of the MVN. Then the updated MVN is then used to obtain the next generation of samples, and the process continues iteratively until convergence or other predefined stopping criteria are met. In our case, the loss function is the sum of square errors (SSE) between values obtained from density functional theory calculations and molecular dynamics simulations. The algorithm converged after $1.3 \times 10^4$ iterations as shown in Figure 2, thanks to the use of the step-wise optimization technique already employed by Wang et al. [77]. Indeed, similar to the study conducted by Wang et al. [77], we capitalized on the physical-inspired structure of the ReaxFF force field, where parameters are categorized based on interaction and atom type, as outlined in Table 2. The approach consisted in gradually optimizing a selection of parameters relevant to particular types of interactions, instead than letting the algorithm operate in a very high-dimensional search space. Hence, we proceeded by optimizing initially the values associated with the bond interactions, then the values related to the dispersion energy, as evidenced by the leap in the loss function in Figure 2. However, after these first two subsets, further optimization of angle and torsional interactions was not feasible because all the initial guess values for their coefficients produced similar loss functions, which prevented the calculation of the second generation of parameters and covariance matrix for the CMA-ES algorithm. More information about the full protocol is provided in the methods section.

**Table 2** Listing of ReaxFF coefficients for each section in the force field file, organized by interaction type. Please note that the numbers in the second column represent the number of coefficients per single entry (e.g., for a single atom type or a single bond). However, the total number of parameters required in the force field depends on the total number of atom types ($N_A$), as indicated in the third column.

| Type | N. of parameters | Note |
|---|---|---|
| General | 41 | |
| Atoms | 32 | for each atom type $N_A$ |
| Bonds | 16 | for each bond $N_{Bond} = \dfrac{(N_A + 1)!}{(N_A - 1)!2!}$ |
| Off-diagonal | 6 | for each heterogeneous pair $N_{vdW} = \dfrac{N_A!}{(N_A - 2)!2!}$ |
| Angles | 7 | for each possible angle in the system $N_{Ang} \leq \dfrac{(N_A + 2)!}{(N_A - 1)!3!}$ |
| Dihedral | 7 | for each possible dihedral angle in the system $N_{Tors} \leq \dfrac{(N_A + 3)!}{(N_A - 1)!4!}$ |
| Hydrogen bonds | 4 | usually only for O, C and N atoms |

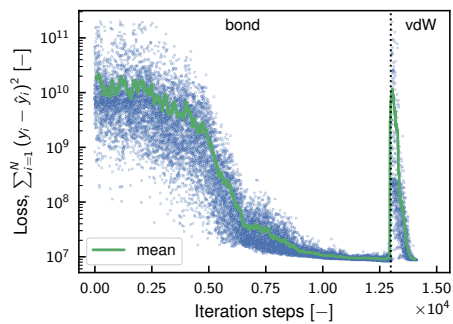

**Fig. 2** Evolution of the loss function during optimization of parameters. The loss function at each iteration is shown as blue points, with the green line representing the moving average of the loss function using a window of 100 iterations. The vertical dashed line indicates the change in the subset of parameters being optimized, i.e., the bond interaction (bond) and the van der Waals interaction (vdW).

The new parameterization has significantly improved the ability to describe the solid state of the inorganic salt, as demonstrated by the Figure 3. The figure includes an energy-strain curve (Figure 3a) and Murnaghan equation of state (EOS) (Figure 3b). The first plot was obtained by applying a shear strain $\varepsilon_{xy}$ to a $2 \times 2 \times 2$ supercell of the stable LiF crystal. While the EOS was computed by expanding and contracting the system using a three-dimensional volumetric strain, since the two are connected by the following relation $\Delta V/V = \varepsilon_{xx} + \varepsilon_{yy} + \varepsilon_{zz}$. At a glance, it is clear that the previous parameterizations were unable to predict the condensed state of LiF accurately. Indeed, neither the force field developed by Yun et al. [42], nor that of Wang et al. [44] showed the presence of a minimum configuration in both the elastic strain and EOS cases. The shear deformation case was particularly problematic because the inverted concavity of the curves led to a nonphysical negative value for the elastic tensor component $c_{12}$, where the stress tensor is defined as $c_{ij} = \sigma_{ij}\varepsilon_{ij}^{-1}$ with $\varepsilon_{ij}$ representing the strain tensor and $\sigma_{ij}$ the stress tensor. The concavity of the energy-strain plot determines this component since it is the function of the second partial derivative of the energy with respect to strain, i.e., $c_{12} = V_0^{-1} \cdot \partial^2 E/\partial\varepsilon_1 \partial\varepsilon_2$. On the other side, thanks to the newly designed database, the reoptimized ReaxFF force field accurately predicts equilibrium configurations for both EOS and elastic strain curves. Specifically, the prediction of the shear strain closely matches *ab initio* results in the region close to the equilibrium value, but deviates substantially for high shear strain values, indicating its limitations in predicting material stiffness accurately. Similarly, the ReaxFF prediction of the equation of state agrees well with the reference values, even for significant volume changes. This remarkable improvement of the proposed reparameterization over previous ones is also demonstrated for the one-dimensional deformation of the crystal and for the metastable crystals (as shown in supplementary Figure S18). It is important to note, however, that the new force field predictions are less accurate for metastable cases with hexagonal and body-centered cubic lattice (Figure 1), suggesting that the functionals used to describe various interactions may have limitations in generalizing the energies from the DFT simulations.

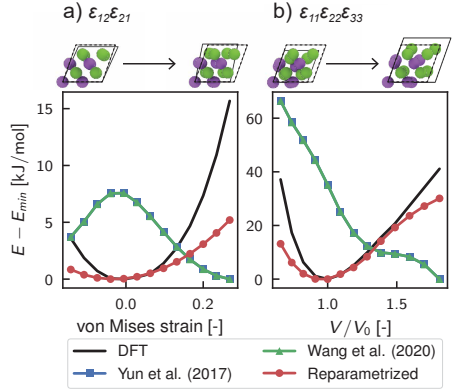

**Fig. 3** Exploring the stability of LiF under stress: A dual analysis reveals the material's response to mechanical manipulation. In (a), the energy-strain plot displays the total energy changes with applied shear strain $\varepsilon_{12} = \varepsilon_{21}$, while (b) depicts the equation of state, i.e., the energy variations versus volume deformation ratio $V/V_0$ ($V_0$ is the volume at the equilibrium). The black line represents DFT results used for training, while the blue, green, and red lines illustrate the energy predictions made by the ReaxFF from Yun et al. [42], Wang et al. [44], and the proposed new reparameterization, respectively. The simulation snapshot above each plot illustrates the frontal view of the crystal at two extreme deformation cases, with the dashed black line indicating the minimal energy unit cell. Notice that green and blue lines are mostly overlapped in the reported pictures.

## ReaxFF prediction of bulk LiF

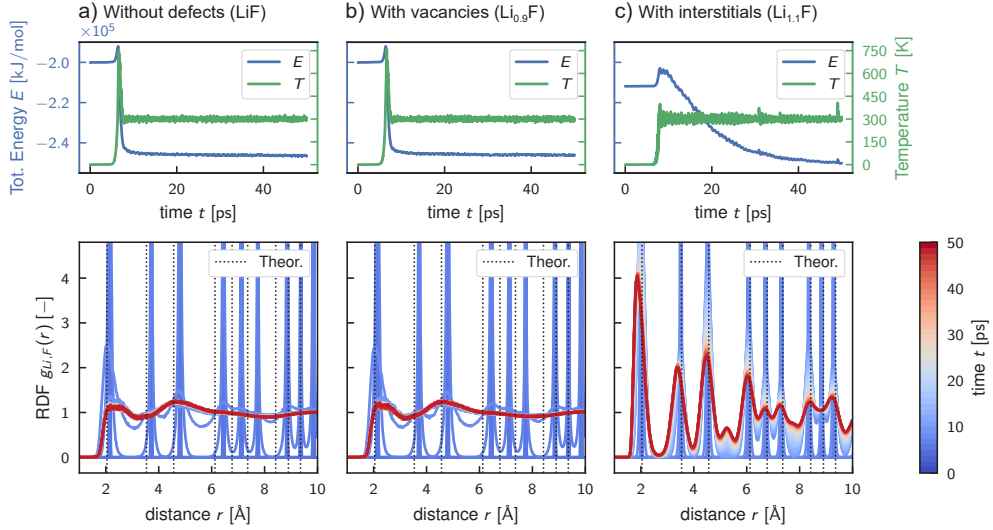

**Fig. 4** The evolution of bulk LiF simulations using reactive molecular dynamics at 300 K using three ReaxFF: Yun et al. [42] (a), Wang et al. [44] (b), and the new reparameterization (c). The top graphs show the total energy trends in a solid blue line and the temperature trends in a solid green line. The bottom panels, on the other hand, show the time evolution of the Li-F radial distribution function (RDF). Each RDF curve is obtained by averaging the trajectory every 1.25 ps and then coloured according to the simulation time indicated by the colour bar on the right. Remarkably, RDF curves from literature force field resemble more systems in a liquid phase rather than solid phase.

To validate the new parameterization, as first test case, we conducted reactive MD simulations using all the discussed ReaxFFs for systems with Li and F atoms to simulate bulk LiF at ambient conditions (300 K, 1 bar). The simulation was performed on a relatively big system with respect to the ones used for the training. Indeed, a supercell of the LiF crystal created by replicating the conventional unit cell five times in all directions was placed in a simulation box of size $20.4 \times 20.4 \times 20.4;\text{Å}^3$, which is analogous to the system in Figure 1d. After the initial relaxation of the system, including box size adjustment, the pure LiF crystal was simulated for 50 ps under an NVT ensemble (constant number of atoms, volume, and temperature) set at 300 K and with a time-step of 0.25 fs. In Figure 4 we report the total energy, temperature, and radial distribution function (RDF) between the Li and F atoms, $g_{Li,F}(r)$, evaluated during the reactive MD simulations with different potentials. Comparing the results from the simulations using the Yun et al. [42] and Wang et al. [44] force fields (Figure 4a and Figure 4b), we observed similar behavior, indicating minimal differences in the parameters for Li and F used in both ReaxFF. Furthermore, we can observe that both force fields failed to accurately describe the simulated system, as the thermal agitation overpowered the binding energy, resulting in the amorphization of the inorganic salt when it reached the external heat bath temperature. This resulted in a sudden drop in the total energy at 8 ps and the increase in temperature upto 750 K, as a consequence of the rapid conversion of potential energy into kinetic energy that results into an artificial phase change. Indeed, since that the melting temperature of LiF is 1121.35 K (848.2 °C) [78], the system should ideally exist in a solid phase in the simulated virtual conditions. This remarkable and unexpected result is even more evident looking at the two radial distribution functions in Figure 4a and Figure 4b, where the initial peaks due to the ordered structure of the system and immobility of the atoms disappear, and the curves become liquid-like RDF as the temperature rises. Moreover, we can notice a shift in the initial RDF peaks with respect to the theoretical position represented by the vertical dotted line, indicating an anomaly in the LiF solid state description of the two FFs. In contrast, our new parameterization performed well in simulating a simple bulk LiF system. Indeed, Figure 4c shows that the radial distribution function has initial peaks that align with theoretical values and remain consistent even at high temperatures. In this case, the thermal agitation of the atoms resulted only in the smoothening of the main peaks, and the emergence of secondary peaks due to the oscillation of some atoms between the lattice and interstitial positions. This important improvement is also evident by visualizing the temperature growth, which gradually reaches 300 K and is held constant as the system converges to an equilibrium state, as visible from the total energy (Figure 4c).

## Diffusion of Li in LiF prediction from ReaxFF

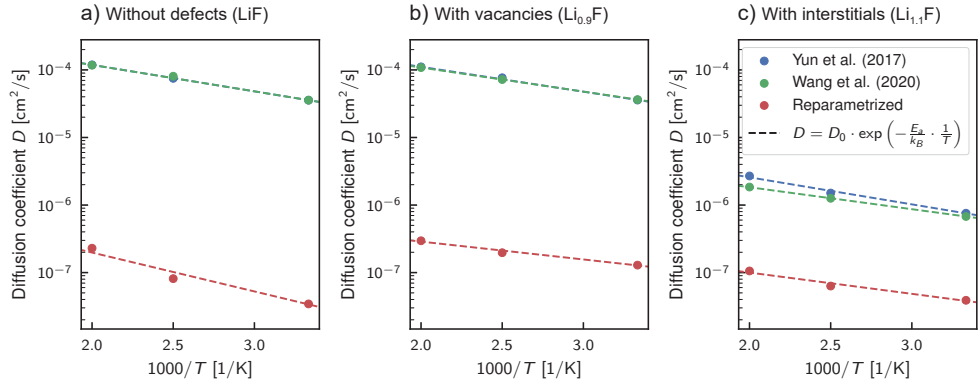

**Fig. 5** Arrhenius plot displaying the diffusion coefficient of Li ($D$) in pure bulk LiF (LiF) and with point defects of vacancy ($Li_{0.9}F$) and interstitial ($Li_{1.1}F$) types, obtained through the molecular dynamics simulations. The blue, green, and red dots represent the $D$ values obtained using the Yun et al. [42], Wang et al. [44], and the proposed new ReaxFF parameterization, respectively. The drawn dashed line results from a linear regression of the coefficients using the least square method.

In designing the training set, we aimed at enhancing the accuracy of predicting both the mechanical and thermophysical properties of the inorganic material through ReaxFF simulation. To achieve this, we included configurations with point defects like vacancies, interstitials, and substitutions and incorporated data from *ab initio* simulations conducted at different temperatures. Therefore, the second test case for the new ReaxFF was to compute the lithium transport properties with a campaign of MD simulations. We evaluated the mass diffusivity of Li atoms in three systems illustrated on the right side of Figure 1, specifically: one with 10 % interstitial lithium atoms ($Li_{1.1}F$), one with 10 % lithium vacancies ($Li_{0.9}F$), and one with no defects (LiF). These systems were simulated for 500 ps at three different NVT temperatures (300 K, 400 K, and 500 K). The lithium diffusion coefficients were extracted by measuring the mean square displacement (MSD) during the simulation and by using the Einstein-Smoluchowski diffusion equation [79–81]:

$$\lim_{t \to \infty} \langle \|\mathbf{r}_i(t) - \mathbf{r}_i(0)\|^2 \rangle_{i \in Li} = c_d D t, \qquad (1)$$

where the limit argument is the MSD definition, $t$ is time, $D$ is the diffusion coefficient, and $c_d$ is a constant indicating dimensionality, with values of 2 for one-dimensional, 4 for two-dimensional, and 6 for three-dimensional diffusion. We obtained the diffusion coefficient values by determining the slope of the straight line that best fits the MSD evolution for each simulation (as seen in the supplementary Figure S19-S21). In Figure 5, we plotted them on a logarithmic scale and used the inverse of the temperature as abscissa. This enabled us to extract the activation energy that appears in the Arrhenius equation:

$$D = D_0 \cdot \exp\left(-\frac{E_a}{k_B T}\right). \qquad (2)$$

Here, $D$ is the lithium diffusion coefficient from the MD simulations, $D_0$ is the maximum diffusivity value (the limit value at infinite temperatures), $k_B$ is the Boltzmann constant, and $E_a$ is the activation energy. We repeated this process for all three ReaxFFs discussed, and the resulting values of $D_0$ and $E_a$ are reported in Table 3. The new force field provides an improved description of bonding forces, which results in a significant reduction in the mobility of lithium atoms within the crystal lattice, as expected. For example, in the case study of defect-free LiF supercell at 300 K, the diffusion coefficient predicted by both the Yun et al. [42] and Wang et al. [44] force fields is $3.56 \times 10^{-5}\,cm^2/s$, which is significantly closer to the diffusivity of lithium ions in the electrolyte ($0.5 \times 10^{-5} - 1.4 \times 10^{-5}\,cm^2/s$ [82]) rather than in the solid. In contrast, the new parameterization predicts a diffusivity of $3.44 \times 10^{-8}\,cm^2/s$, which is a reduction of 3 orders of magnitude. Similar reductions in diffusivity are seen in cases involving interstitials and vacancies.

We compared the obtained values with *ab initio* values calculated by Zheng et al. [29]. They studied lithium diffusion in various organic components of the SEI by exploring the energy surface using the surface energy Climbing Image Nudged Elastic Band (CI-NEB) method [83]. The exploration of the energy surface by Zheng et al. confirmed that the diffusion of lithium in lithium fluoride occurs via three distinct possible mechanisms. These mechanisms include: vacancy movement, where lithium atoms jump to the adjacent empty lattice site (vacancy); direct-hopping, where lithium moves directly from one interstitial site to another; and the knock-off mechanism, in which an interstitial lithium atom replaces a lithium atom in the crystal lattice, thus causing the displaced lithium atom to move into another interstitial site. Figure 6 shows the Arrhenius curves for the three diffusion mechanisms calculated using the values of activation energy and maximum diffusion obtained from the *ab initio* study, as well as the curves obtained from the reactive molecular dynamics simulations. Regarding the maximum diffusion coefficient $D_0 = 3\alpha^2 v$, Zheng et al. approximate its value using the estimated phonon frequency in the crystal $v = 1 \times 10^{13}\,\text{s}^{-1}$ and the migration distance for the lithium ions $\alpha$ observed during the CI-NEB simulations [29]. Molecular dynamics simulations can exhibit various diffusion mechanisms, leading to the calculated curves for the new force field and each studied system falling within the region defined by the three diffusion mechanisms. For the calculated temperature values, the diffusion coefficients turn out to be intermediate values between pure knock-off diffusion and pure direct-hopping diffusion. However, looking at the slope of the Arrhenius curves predicted by molecular dynamics simulations in Figure 6 it is clear that even the new ReaxFF underestimated the activation energy. Indeed, even for the most favorable diffusion mechanism (knock-off), the activation energy obtained from the *ab initio* energy profile is $24.1\,\text{kJ mol}^{-1}$, which is more than double the value calculated with the new force field for the defect-free case ($11.0\,\text{kJ mol}^{-1}$). The discrepancies are even more significant for all other cases listed in Table 3.

**Table 3** The values of the activation energy, $E_a$, and the maximum value of the lithium self-diffusion coefficient, $D_0$, for the three simulated systems and for each ReaxFF used. The errors were estimated based on the inference for the linear regression model coefficients (least squares method), and assuming a 95 % confidence interval.

| | **ReaxFF** | $\mathbf{D_0}$ [cm$^2$/s] | $\mathbf{E_a}$ [kJ mol$^{-1}$] |
|---|---|---|---|
| LiF | Yun et al. | $(7.23 \pm 0.05) \times 10^{-4}$ | $7.50 \pm 0.02$ |
| LiF | Wang et al. | $(7.2 \pm 0.5) \times 10^{-4}$ | $7.0 \pm 0.2$ |
| LiF | This work | $(3 \pm 2) \times 10^{-6}$ | $11.0 \pm 1.6$ |
| Li$_{0.9}$F | Yun et al. | $(6.1 \pm 0.3) \times 10^{-4}$ | $7.1 \pm 1.4$ |
| Li$_{0.9}$F | Wang et al. | $(5.5 \pm 0.2) \times 10^{-4}$ | $6.810 \pm 0.013$ |
| Li$_{0.9}$F | This work | $(9.0 \pm 1.7) \times 10^{-7}$ | $5.1 \pm 0.5$ |
| Li$_{1.1}$F | Yun et al. | $(1.5 \pm 0.3) \times 10^{-5}$ | $7.6 \pm 0.5$ |
| Li$_{1.1}$F | Wang et al. | $(8.2 \pm 1.3) \times 10^{-6}$ | $6.22 \pm 0.04$ |
| Li$_{1.1}$F | This work | $(4.0 \pm 1.2) \times 10^{-7}$ | $6.0 \pm 0.8$ |

To address this discrepancy, we conducted further analysis of the ReaxFF by reproducing the energy curves for lithium diffusion by vacancies and by direct-hopping using the CI-NEB method. Initial and final configurations were made using a $2 \times 2 \times 2$ LiF supercell and inserting the lithium atoms into two adjacent sites identified with the Voronoi analysis from the pymatgen python library [50]. While for the vacancy diffusion case, one lithium was removed from the crystal lattice from two adjacent unit cells. The energy profile was then obtained using the CI-NEB algorithm with the same parameters used in the DFT simulation for training the new ReaxFF. To ensure convergence, we used 23 images for direct-hopping diffusion and 17 images for vacancy diffusion. Our simulations yielded activation energies of $64.3\,\text{kJ mol}^{-1}$ for vacancy diffusion and $87.5\,\text{kJ}\,\text{mol}^{-1}$ for direct-hopping. The value for vacancy diffusion was comparable to previous work ($63.7\,\text{kJ}\,\text{mol}^{-1}$). Still, the value for direct-hopping is higher than that calculated by Zheng et al. ($52.1\,\text{kJ}\,\text{mol}^{-1}$) due to the use of a smaller system to reduce the computational cost, resulting in a higher defect density in our case. Subsequently, from the result obtained from the DFT simulations, we proceeded to calculate each image's energy using the various ReaxFF discussed. In Figure 7 we show the comparison of the energy profile predicted by the various methods.

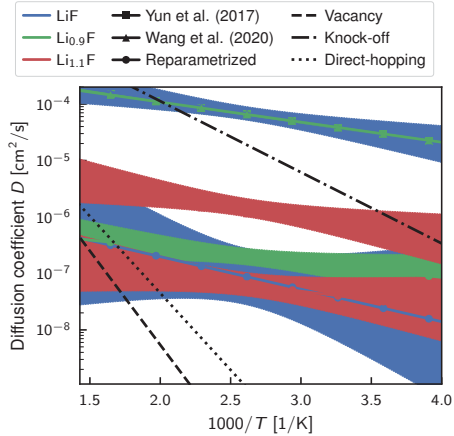

**Fig. 6** Arrhenius plot of diffusion comparing the predictions from Reactive MD and DFT simulations. The interpolation from MD results is shown using continuous lines and markers (square for Yun et al. [42], triangular for Wang et al. [44], and circular for the new reparameterization), and the confidence interval of 95 % is shown with the shaded areas. The effect of defects on diffusion is indicated by different colors: blue color for the defects-free case, green for the cases with vacations, and red for the case with interstitial lithium. The black dashed, dash-dot, and dotted lines represent the Arrhenius curves estimated by Zheng et al. [29] for three potential transport mechanisms (vacancy, knock-off, and direct-hopping) from Climbing Image Nudged Elastic Band (CI-NEB) studies.

The additional investigation of ReaxFF presented in Figure 7 reveals significant discrepancies between previous parameterizations, DFT prediction, and the reoptimized force field. Specifically, the energy barrier value prediction was found to be incorrect for all the ReaxFF studied, and the new force field exhibited unrealistic behavior in the case of interstitial diffusion. Indeed, the maximum energy values corresponded to the initial and final configurations, while the minimum value corresponded to the transition configuration. This behavior contradicts the physical nature of the phenomena, and the predictions made by the DFT calculation and even by previous parameterizations by Yun et al. [42] and Wang et al. [44]. In order to be able to explain this numerical artifact, it is important to consider the general structure of ReaxFF, which

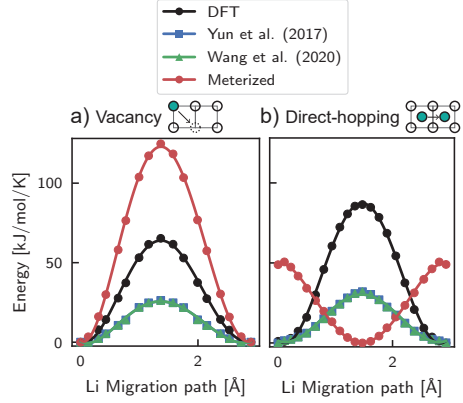

**Fig. 7** Graphical visualization of the energy profile for lithium migration by vacancy (a) and direct-hopping (b) calculated by CI-NEB is depicted by the black line and circular markers. The markers indicate the exact energy value of each distinct image obtained from the convergence of the CI-NEB algorithm and were used to calculate the energy using three different methods: ReaxFFF by Yun et al. [42], Wang et al. [44], and a new reparameterization, represented by blue, green, and red colored markers and line, respectively.

is described by the provided equation:

$$V_{reaxFF} = \underbrace{V_{bond} + V_{angle} + V_{dihedral}}_{\text{bonded interaction}} \\ + \underbrace{V_{over} + V_{under}}_{\text{bonded interaction (coordination)}} + \underbrace{V_{vdW} + V_q}_{\text{non-bonded interaction}} \qquad (3) \\ + \underbrace{V_{conj} + V_{trip} + V_{C_2} + V_{H-bond} + V_{lp}}_{\text{system specific}} .$$

$V_{reaxFF}$ represents the total reactive potential of the particle $i$ interacting with neighboring atoms, which can be divided into three contributions: bonded, non-bonded, and "system specific". The first describes the bonded interaction between atoms and is formed by the bond energy $V_{bond}$, the angle energy $V_{angle}$ and the dihedral (or torsional) energy $V_{dihedral}$ and the corrective terms over-coordination $V_{over}$, and under-coordination $V_{under}$ energies. The non-bonded energy, which is analogous to classical molecular dynamics, is given by the sum of the van der Waals potential $V_{vdW}$ and Coulomb potential $V_q$. The last terms, on the other hand, are a set of corrective energies that accounts for specific phenomena in certain specific systems. For additional details about the mathematical and physical implications of each term, the reader can refer to the works by Van Duin et al. [37] and Chenoweth et al. [84].

From Eq. (3) and the previous studies, we can attribute the unrealistic behavior observed with the new force field to the increased energy contribution from bond energy, which improves the description of system connectivity. However, this results in an increase in the effect of correction energy terms, specifically over-coordination energy $V_{over}$, and under-coordination energy, $V_{under}$, in Eq. (3), which are functions of the difference between the total number of bonds of an atom and its valence number [37]. Indeed, in the initial and final conditions, the lithium atom has the same number of neighboring atoms, which is higher than in the transition state, where space is created by moving the lattice atoms, and the increase of the lithium distance respects the atoms at the extremes of the unit cell. This unexpected artifact highlights the critical importance of carefully considering energy contributions in force fields to ensure accurate predictions, as observed in the case study using ReaxFF. Consequently, it is likely that additional configurations are needed to be included in the training set. However, it is important to note that this alone does not guarantee the resolution of all the aforementioned issues. In fact, due to the reliance on a fixed functional shape in the force field, a more deep and sophisticated alteration in the mathematical expression of the energy terms might be required. Implementing such changes would involve modifying the code, which exceeds the scope of this work. This also highlights that a critical challenge here is the limited transferability of the parameterization beyond the specific database, emphasizing the necessity of a comprehensive database of *ab initio* simulations that is specifically tailored and adequately representative for studying the targeted system at an atomistic scale.

# Conclusion

In conclusion, the proposed partial reparameterization methodology for the ReaxFF has significantly improved the description of lithium fluoride in the solid state, leading to better predictions of its solid phase properties and lithium mobility in LiF crystal using reactive MD simulations. Indeed, implementing the new parameters of the ReaxFF has demonstrated its ability to predict the stable unit cell under mechanical deformation accurately, exhibit typical solid-state RDF, and notably reduce lithium diffusivity in LiF by at least two orders of magnitude. The automation and interactivity of the protocol, achieved by leveraging Python libraries for atomistic simulations, made it possible to construct and simulate various configurations of the LiF crystal needed to build a database for correcting bond and van der Waals interactions of the ReaxFF. The new force field obtained from the reoptimization not only improved the behavior of the crystal in the solid phase but also partially corrected the description of lithium transport phenomena in LiF.

However, the in-depth investigation carried out revealed that the diffusion activation energies predicted by the new force field are still underestimated. This limitation may be due to the method used to construct the database, which did not directly sample lithium transport, but focused more on local or global deformation of the crystal lattice. This study highlights the strong dependence of the ReaxFF on the configurations included in the database and the challenges in adequately interpolating the energy surface in unexplored regions. The need to correct the ReaxFF for each specific case makes it challenging to adopt it for the up-scaling electronic simulations (here DFT) up to the molecular level. Based on our experience, the need for ever-larger databases poses serious questions on the cost-benefit ratio of this method. Although the ReaxFF MD simulations are much faster than *ab initio* simulations, given all the functionals to be computed and the need to update the charges at each step reduces by at least a factor of 100, the speed of MD simulations compared to simulations with Lennard-Jones fluid [85].

In addition to the sensitivity of the ReaxFF to the training set used, the difficulty in accurately describing the system studied in this work may also be attributed to an intrinsic bias of the force field. Indeed, it is worth noting that the original formulation of the ReaxFF potentials proposed by Van Duin in 2001 was validated primarily on organic systems [37], and various new functionals have been added to improve the accuracy and applicability of ReaxFF in other systems over the past few decades [86–88]. These include the "lone-pair" energy term ($V_{lp}$ in Eq. (3)) for hydrocarbon combustion [84, 86, 89], a three-body functional ($V_{trip}$ in Eq. (3)) term for $-NO_2$ group chemistry [86, 90, 91], and an energy term ($V_{H-bond}$ in Eq. (3)) to account for hydrogen bonds in aqueous systems [84, 92], among others. Therefore, it may be necessary to introduce new corrective energy terms to improve the accuracy of ReaxFF in strongly inorganic systems such as LiF.

On the other hand, the rapid development of neural network-based force fields [93] (NNFF) may provide alternative and accurate approaches to the ReaxFF. The database constructed using the proposed protocol could be used for training these new force fields, which can provide superior performance at the cost of compromising on the physical insight into the parameters obtained from the training. In summary, the proposed methodology can be extended to the parameterization of other potentials, and by increasing the number of initial configurations, it may also be possible to proceed with the parameterization of neural network potentials. Moreover, the guided reparametrization method could incorporate other frequently encountered inorganic compounds in the SEI within future ReaxFF or NNFF. This expansion offers an exciting opportunity to explore increasingly intricate and realistic systems resembling mosaic structures [94] that emulate the Peled model [7].

# Methods

## Interactive reparameterization protocol

The protocol for calculating the new ReaxFF parameters is presented in a flowchart shown in Figure 8. The procedure is carried out using four Jupyter Notebooks (JNBs) [54, 55] that facilitate the automatic construction, visualization, and simulation of the necessary configurations for database construction and optimization of the reactive potential. Python libraries ASE [48, 49], pymatgen [50, 51], PLAMS [69, 70] and ParAMS [52, 53], designed for atomistic systems manipulation and simulation are utilized throughout the entire process, allowing for streamlined and efficient handling of the various steps.

The first JNB initiates the protocol by defining the chemical formula of the component to study. Then, using the API from the Materials Project database, the available crystalline units are downloaded. In the case of LiF, there are three configurations available, corresponding to different lattices (Figure 1.a-c), one stable (Fm$\bar{3}$m) and two metastable (P6$_3$mc, Pm$\bar{3}$m). These crystals are imported into the interactive work environment as virtual objects from the ASE (`Atoms`) and pymatgen (`Structure`) libraries. Within this virtual environment, the crystals were manipulated to generate

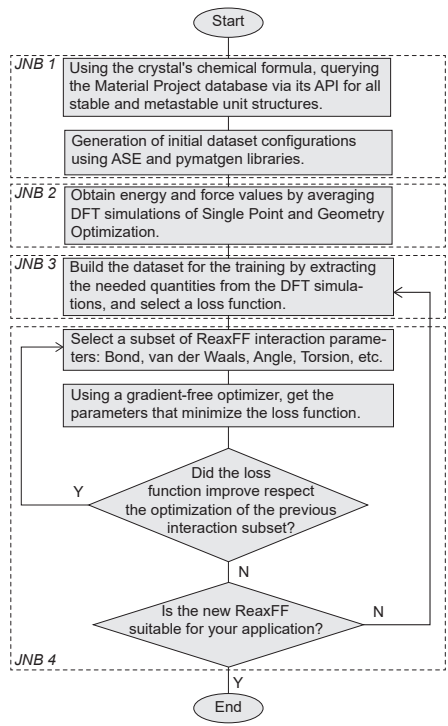

**Fig. 8** Flowchart followed for the reparameterization of ReaxFF. The dotted boxes indicate the Jupyter Notebook (JNB) specific to that part of the workflow, all available at https://doi.org/10.5281/zenodo.8036775 and in the repository https://github.com/paolodeangelis/Enhancing_ReaxFF.

initial configurations for the DFT simulations. Specifically, we build different defected or deformed crystals following the procedure by LaBrosse et al. [73], namely:

***Supercells:*** To capture the effect of the atoms' coordination in the training set, several supercells of size $2\times 1\times 1$, $2\times 2\times 1$, $2\times 2\times 2$, $3\times 2\times 2$, $3\times 3\times 2$, and $3\times 3\times 3$ were constructed.

***Vacancies:*** We randomly removed a lithium or fluorine atom from the $3\times 2\times 2$ supercell, up to a maximum of five vacancies.

***Strain:*** We applied different types of strain to each LiF crystal, including a normal strain $\varepsilon_{11}$, a shear strain $\varepsilon_{12} = \varepsilon_{21}$, and a homogeneous deformation in all directions $\varepsilon_{11} = \varepsilon_{22} = \varepsilon_{33}$ (needed for computing the equation of state), resulting in 13 configurations with strain values ranging from $-12.5\%$ to $23.5\%$.

***Substitution:*** We randomly substituted a lithium or fluorine atom with the opposite species, with a maximum of five possible defects for each crystal.

***Interstitial:*** We inserted an interstitial atom at the center of Voronoi volumes obtained from the $3\times 2\times 2$ supercell for each LiF crystal. We repeated this procedure five times to create structures with 1, 2, ..., and 5 interstitial atoms.

***Slab:*** To include surface energy in the ReaxFF, we generated crystal slabs for the surfaces (100), (110), (111), and (210). We repeated each slab 2, 3, or 4 times to account for different thicknesses and we introduced sufficient empty space along the normal direction to avoid numerical artifacts due to the periodic boundary conditions (PBCs).

**Bulk at 300 K and 500 K**: We included frames of *ab initio* MD simulations at 300 K and 500 K for the $3 \times 2 \times 2$ supercell to predict possible energy fluctuations during ReaxFF MD.

**Amorphous LiF**: To account for the high-energy state of the crystal, we included possible amorphous LiF with an *ab initio* MD simulation at $T = 2500$ K.

In the second JNB, previously constructed configurations are used to perform *ab initio* simulations to obtain the numerical values needed for the databases. The ASE and pymatgen objects of the various systems were used as input to the PLAMS library of the AMS [71] commercial code, enabling the running, control, and processing of DFT simulations. For LiF, over 300 DFT simulations were carried out using the BAND plane-wave DFT code available in AMS, as listed in the supplementary Table S1. Single-point (SP) simulations were used to calculate the energy and force values of each atom in each configuration, by solving the Kohn-Sham equations while considering the atom cores as fixed. For systems with defects, geometry optimization (GO) simulations were initially performed to determine the equilibrium configuration. *Ab initio* MD simulations were performed using Grimme's extended density functional based on tight-binding (DFTB) [95] due to their low numerical cost. From the resulting trajectories, 10 frames were selected, and their force and energy values were refined using SP-DFT simulations.

The third and fourth JNBs heavily rely on the AMS ParAMS [52, 53] library, which is specifically designed for potential optimization. In the third JNB, all quantities, such as energy, forces, and charges needed for the database, are extracted, resulting in more than 3000 entries. Regarding the potential energy, because it is a state function, the database does not include the absolute values obtained from the DFT simulations but the relative values obtained with respect to the defect-free configuration of the LiF crystal. This database is then used to optimize the ReaxFF in the fourth and final JNB. The optimization process begins by selecting the group of parameters that will form the search space, starting with those that significantly influence the behavior of the ReaxFF, such as the bond energy, followed by the van der Waals energy and angular energy terms. Various gradient-free optimization algorithms are available in the ParAMS library, and for the ReaxFF optimization for LiF, we chose the genetic algorithm Covariance Matrix Adaptation Evolution Strategy algorithm (CMA-ES) [76] to minimize the objective function represented by the sum of squared errors (SSE) between the DFT values and the ReaxFF values.

Each group or subset of parameters is optimized sequentially, and the ReaxFF optimization process is considered complete when further improvements in the loss function become negligible.

### Density functional theory calculations

The dataset values (energies, forces, charges, etc.) for the ReaxFF optimization were obtained from DFT simulations performed with the commercial BAND [96, 97] code available in the AMS suite. We numerically solved the Kohn-Sham equations using the Perdew-Burke-Ernzerhof (PBE) [98] functional and the polarized double zeta (DZ) numerical atomic orbitals (NAOs) basis set for the calculation of the $s$, $p$, and $d$ orbitals. The software automatically chose the values of k-point and frozen electrons depending on the desired accuracy, and we selected a high accuracy value that guaranteed an error of less than 0.01 eV per atom, and by comparing the formation energy values obtained for each crystalline unit of LiF studied with the values reported on the Material Project online database (see Figure 1a-c). For very inhomogeneous systems, such as those with a large number of interstitial atoms or surfaces, we calculated forces without frozen atoms and using a single zeta (SZ) type basis set to speed up the calculation of the equilibrium configuration. We then refined the resulting configuration using the settings described above.

For detailed instructions on installing and utilizing the protocol and database repository, please refer to the supplementary material provided.

## Molecular Dynamics calculations

All the reactive molecular dynamics simulations were performed using the open-access code LAMMPS [67] with the ReaxFF package [99]. The initial configuration for the diffusion of Li in bulk LiF was obtained by starting from the primitive unit cell of the stable crystal obtained from the Material Project database. The unit cell was then converted into the conventional unit cell using the pymatgen routine `ConventionalCellTransformation` [50] and replicated six times along all directions to obtain a $6 \times 6 \times 6$ supercell with final dimensions of 24.5 Å $\times$ 24.5 Å $\times$ 24.5 Å. To create the system with vacancies (i.e. $Li_{0.9}F$), 86 randomly selected lithium atoms were removed. While to create the system with interstitial atoms (i.e. $Li_{1.1}F$), 86 lithium atoms were placed inside the LiF supercell using the PACKMOL code [100]. After the initial energy minimization, the system was simulated for 0.5 ns using an NVT ensemble at three different temperatures (300 K, 400 K, and 500 K) with a Nose-Hoover thermostat [101] and a relaxation time of $\tau_T = 25$ fs. The integration time step was set to $\delta t = 0.25$ fs. Thermodynamic properties, including the instantaneous mean square displacement for all lithium atoms, were sampled every 50 simulation steps to compute the diffusivity as discussed later.

## Diffusion energy barrier calculations (DFT)

To determine the energy barrier of Li ion diffusion, we used the climbing image-nudged elastic band (CI-NEB) method [83]. To employ this method, we built a $2 \times 2 \times 2$ supercell of the primitive cell of the stable LiF crystal and placed a Li atom in the interstitial site found with the Voronoi analysis [102] in two adjacent cells to study the diffusion by interstitials. While, for the vacancy case, we removed a Li atom from two adjacent cells of the supercell. After geometry optimization of the initial and final states, we performed the CI-NEB calculation using 23 images for the direct-hopping diffusion case and 17 images for the vacancy case. We set the maximum perpendicular force component for the transitional state to be 2.5 eV Å$^{-1}$ as the climbing convergence criterion. Finally, we obtained the activation energy by averaging the energy differences between the initial and transitional state and the final and transitional state.

## Diffusion energy barrier calculations (MD)

To compute the diffusion coefficient $D$ for each combination of temperature, system, and potential, we employed the Einstein-Smoluchowski diffusion equation [79–81], Eq. (1), that requires the time derivative of the MSD obtained from ReaxFF MD simulations. We obtained the numerical MSD from the MD trajectories and fitted it to a linear model of the form $\langle \mathbf{r}^2 \rangle = \alpha_0 + \alpha_1 \cdot t + \varepsilon$, where $\langle \mathbf{r}^2 \rangle$ represents the mean-square displacement, $\varepsilon$ is the statistical error, and $t$ denotes the elapsed simulation time (Figure S19-S21). To determine the coefficients $\alpha_0$ and $\alpha_1$, we used the least squares method (LSM) [103]. Consequently, the diffusivity can be evaluated as $D = \alpha_1/6$.

An analogous procedure was followed to compute the activation energy from the Arrhenius law, Eq. (2). To linearize the equation, we took the logarithm of both sides and applied the variable substitution $x = T^{-1}$. This manipulation resulted in the equation taking the form $\ln(D) = \ln(D_0) - E_a/k_B \cdot x$ (Figure 5). Using the diffusion coefficients obtained from the ReaxFF MD simulations, we fitted a linear model of the form $\ln(D) = \beta_0 + \beta_1 \cdot x + \varepsilon$ to obtain the activation energy as function slope of the line, i.e., $E_a = -\beta_1 \cdot k_B$.

To obtain the reported 95% confidence intervals for the diffusivity and activation energy in Table 3, we assumed that the statistical error $\varepsilon$ of the linear model follows a Student's t-distribution $\mathcal{T}_\nu$, where $\nu$ represents the degrees of freedom of the distribution [103]. In our case, $\nu$ equals the number of sampled data $n$ minus the constraints of our model, which are the intercept and slope of the model ($\nu = n - 2$). Under these reasonable hypotheses, the uncertainty for the diffusivity, $\delta D$, and activation energy, $\delta E_a$, is estimated as follows [103]:

$$
\begin{aligned}
\delta D &= t_{n-2,\,0.025} \cdot \frac{\sqrt{\frac{1}{n-2} \sum_{i=1}^{n} \left( \langle \mathbf{r}^2 \rangle_i - \alpha_0 - \alpha_1 t_i \right)^2}}{6 \cdot \sqrt{\sum_{j=1}^{n} \left( t_j - \langle t \rangle \right)^2}}, \\
\delta E_a &= t_{n-2,\,0.025} \cdot \frac{\sqrt{\frac{1}{n-2} \sum_{i=1}^{n} \left( \ln(D_i) - \beta_0 - \beta_1 \frac{1}{T_i} \right)^2}}{k_B \cdot \sqrt{\sum_{j=1}^{n} \left( \frac{1}{T_j} - \langle \frac{1}{T} \rangle \right)^2}}.
\end{aligned}
\quad (4)
$$

$t_{n-2,\,0.025}$ is the cuts probability 0.025 from the upper tail of Student's t distribution $\mathcal{T}_{n-2}$, with $n-2$ degrees of freedom. $\alpha_0$, $\alpha_1$, $\beta_0$, and $\beta_1$ are the coefficients of the two linear models determined from the linear regression. $\langle \mathbf{r}^2 \rangle_i$ represents the MSD at the $i$-th time $t_i$ of the simulation, $k_B$ is the Boltzmann constant, and $D_i$ and $T_i$ are the diffusivity and corresponding simulation temperature values, respectively. We use the notation $\langle \cdot \rangle$ to indicate the mean value of the independent sampled variable of each model, i.e., the time, $t$, for the first model, and the reciprocal of the temperature, $T^{-1}$, for the second. The confidence interval for the entire linear model depicted in Figure 6 was instead obtained with the Eq. (5), which provides the uncertainty for the logarithm of the diffusivity, $\ln(D)$, as a function of the reciprocal of the temperature, $T^{-1}$, which is the independent variable of the linear model [103]:

$$\delta\left[\ln(D)\right]\left(\frac{1}{T}\right) = t_{n-2,\,0.025} \cdot \sqrt{\frac{1}{n-2}\sum_{i=1}^{n}\left(\ln(D_i) - \beta_0 - \beta_1\frac{1}{T_i}\right)^2} \cdot \sqrt{\frac{1}{n} + \frac{\left(\frac{1}{T} - \left\langle\frac{1}{T}\right\rangle\right)^2}{\sqrt{\sum_{j=1}^{n}\left(\frac{1}{T_j} - \left\langle\frac{1}{T}\right\rangle\right)^2}}}. \quad (5)$$

## Acknowledgements

The project leading to this application has received funding from the European Union's Horizon 2020 research and innovation programme under grant agreement No 957189 (BIG-MAP project). The authors also acknowledge that the results of this research have been achieved using the DECI resource ARCHER2 based in UK at EPCC with support from the PRACE aisbl.

## Author contributions statement

The research project was conceptualized and designed collaboratively by P.D.A., M.F., P.A., and E.C. In particular, they collectively discussed the feasibility and suitability of using ReaxFF for investigating the SEI formation. E.C. and P.D.A. subsequently decided to focus on the LiF compound. P.D.A. developed, tested, and refined the protocol for reoptimizing the ReaxFF, and conducted the simulations and postprocessing of the data with the assistance of R.C. P.D.A., R.C. and M.F. worked on analyzing and interpreting the results, and P.D.A. wrote the manuscript with input from R.C., M.F., E.C. and P.A. The project was supervised by E.C. and P.A. All the authors have read and approved the final manuscript.

## Additional information

Together with the material presented in this article, there is a supporting information document that contains further plots and data about the training data for the reparameterization and the mew ReaxFF file, as well as an additional plot of Li diffusion analysis. This supporting information document can be found at https://TBD.
**Accession codes**
In this study, we employed several computational tools to carry out our simulations. We used the open-source LAMMPS code for the ReaxFF MD simulations, while the commercial codes BAND and DFTB were utilized for the DFT and ab initio MD simulations, respectively. These last two codes are all available in the Amsterdam

Modeling Suite by SCM. In addition, we employed several Python libraries, including ASE, Pymatgen, and Packmol, to arrange the initial configurations, as well as the PLAMS library and an in-house Python code for automatically managing the simulation campaign. Finally, we utilized the ParAMS library to perform the ReaxFF parametrization.

The repository containing all the Jupyter notebooks required for running the ReaxFF reparametrization process, which extensively utilized the aforementioned codes, can be accessed at https://github.com/paolodeangelis/Enhancing_ReaxFF. Furthermore, the energies and forces obtained from DFT simulation are stored as ASE SQLite3 and can be found in the repository located at https://github.com/paolodeangelis/Enhancing_ReaxFF_DFT_database.

**Competing interests**

The authors declare no competing financial interest.

# Enhancing ReaxFF for Molecular Dynamics Simulations of Lithium-Ion Batteries: An interactive reparameterization protocol (Supplementary Material)

**Paolo De Angelis**[1,*], **Roberta Cappabianca**[1], **Matteo Fasano**[1], **Pietro Asinari**[1,2,†], **and Eliodoro Chiavazzo**[1,‡]

[1]Department of Energy "Galileo Ferraris", Politecnico di Torino, Corso Duca degli Abruzzi 24, 10129 Torino, Italy.
[2]Istituto Nazionale di Ricerca Metrologica, Strada delle Cacce 91, 10135 Torino, Italy.
*paolo.deangelis@polito.it
†pietro.asinari@polito.it
‡eliodoro.chiavazzo@polito.it

This is the supplementary material for the article "Enhancing ReaxFF for Molecular Dynamics Simulations of Lithium-Ion Batteries: An interactive reparameterization protocol." Here we list resources and images that aim to provide further explanation and understanding of the methods and results explained in the main text.

In line with the FAIR principles (Findability, Accessibility, Interoperability, and Reusability)[1], we provide detailed documentation within the two repositories (Section 1 and Section 2) that accompany the article. These repositories are designed to help the reproducibility of our results and facilitate further exploration of the LiF configurations or other Solid Electrolytes Interphase (SEI)[2] compounds, extending the database and improving the ReaxFF force field[3].

## Contents

# 1 Enhancing ReaxFF protocol (Protocol Repository)

In this repository, we collect and organize all the steps described in the main text for reparameterizing the ReaxFF potential. To facilitate the interaction with the protocol, we have divided the workflow, illustrated in Figure S1, into four main Jupyter Notebooks (JNBs) and an auxiliary notebook: `JNB1-Initial_configurations.ipynb`, `JNB2-Simulations.ipynb`, `JNB3-Build_trainingset.ipynb`, `preJNB4-ReaxFF_optimization.ipynb`, `JNB4-ReaxFF_optimization.ipynb`. These provided

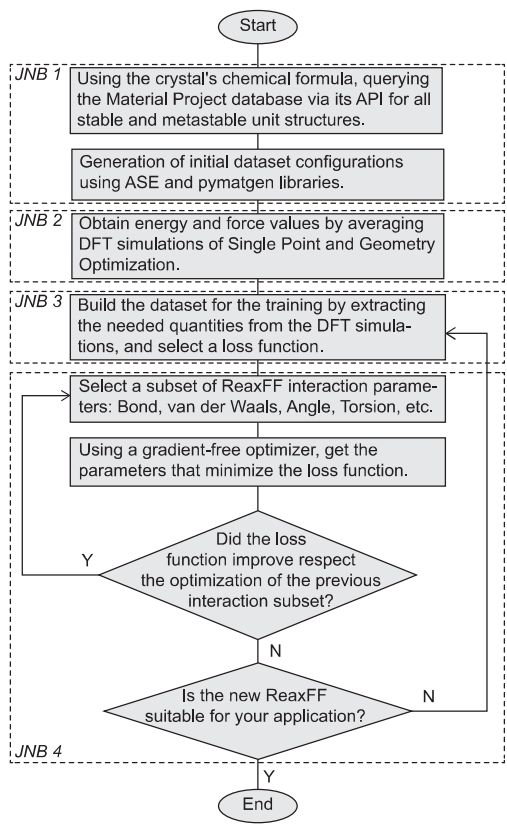

**Figure S1.** Protocol workflow overview

Jupyter Notebooks are specifically designed to streamline the process of configuration building (JNB1), DFT simulation execution (JNB2), database preparation (JNB3), and ReaxFF potential optimization (JNB4), as outlined in the workflow in Figure S1. The conditional part of the workflow is achieved using the fourth and auxiliary notebook since the auxiliary notebook (`preJNB4-ReaxFF_optimization.ipynb`) is selected the subset of ReaxFF parameters related to a specific interaction and then fed to the JNB4, which will change it to minimize the loss function.

The repository is available both in Zenodo at the permanent link https://doi.org/10.5281/zenodo.8036775, and on version control repositories host GitHub https://github.com/paolodeangelis/Enhancing_ReaxFF, Figure S2.

## 1.1 Installation

The protocol strongly relies on the commercial code Amsterdam Modeling Suite (AMS) by Software for Chemistry & Materials (SCM)[4]. However, depending on which part of the protocol you aim to reproduce, it is possible to install only the minimal requirements, as described in Subsubsection 1.1.1, which allow for the generation and handling of atomistic simulations. To perform the simulations, additional requirements outlined in Subsubsection 1.1.2 are necessary. For the ReaxFF optimization, the Python library ParAMS[5] is indispensable, and it is currently available

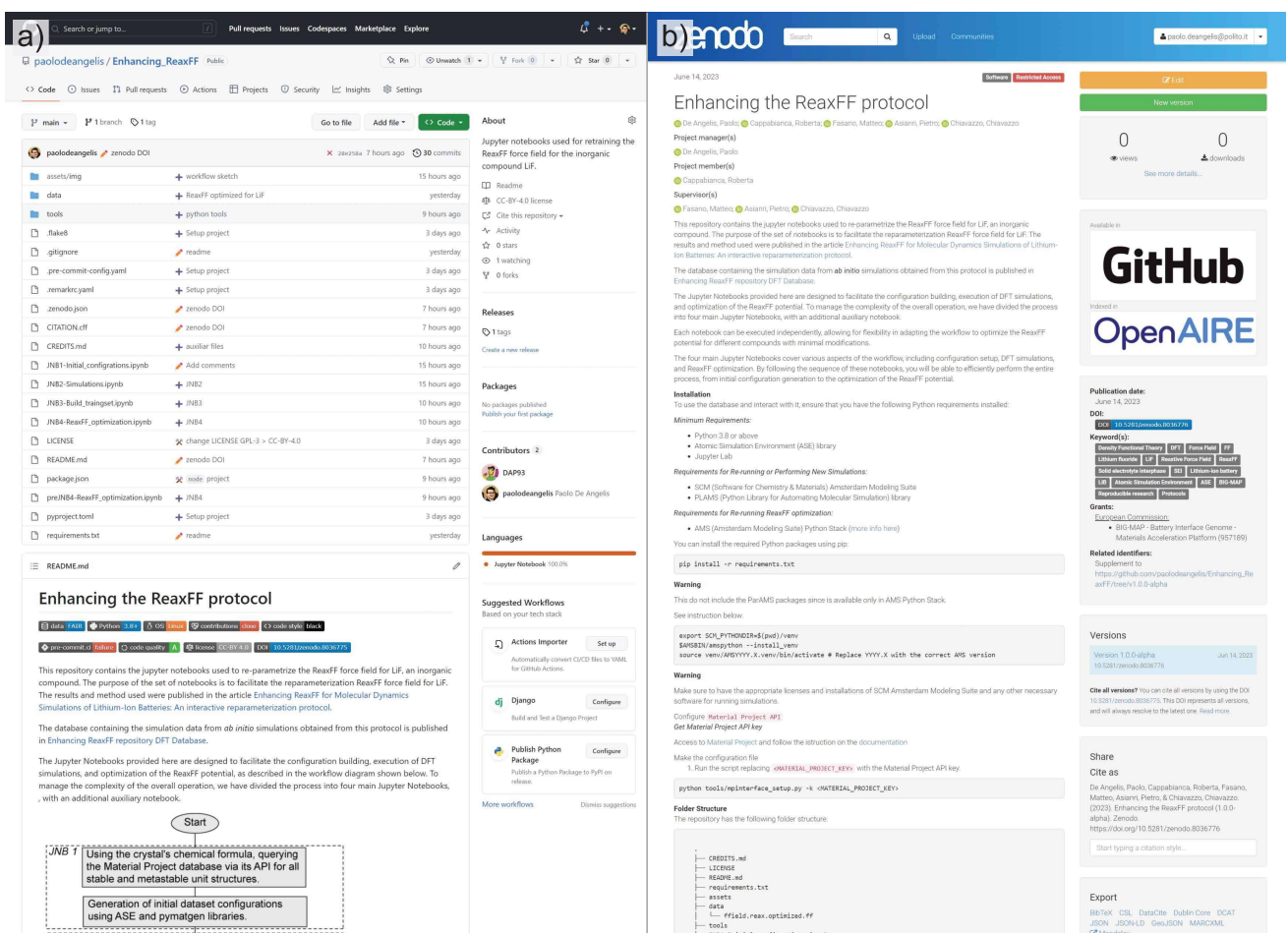

**Figure S2.** Screenshot of the protocol hosted on GitHub (a) and Zenodo (b).

only within the AMS Python Stack, as explained in Subsubsection 1.1.4.

### 1.1.1 Minimum Requirements
- Python 3.8 or above
- Atomic Simulation Environment (ASE) library
- Jupyter Lab

### 1.1.2 Requirements for Re-running or Performing New Simulations
- SCM (Software for Chemistry & Materials) Amsterdam Modeling Suite
- PLAMS (Python Library for Automating Molecular Simulation) library

### 1.1.3 Requirements for Re-running ReaxFF optimization
- AMS (Amsterdam Modeling Suite) Python Stack (more info here). To utilize the AMS Python Stack for this project, we recommend creating a Python virtual environment as follow:

```
export SCM_PYTHONDIR=$(pwd)/venv
$AMSBIN/amspython --install_venv
source venv/AMSYYYY.X.venv/bin/activate # Replace YYYY.X with the correct AMS
    version
```

### 1.1.4 Python project setup

Regardless of the software available, the next step is to install all the additional Python packages (such as pymatgen, ASE, etc.) required for the protocol. This can be accomplished using the Python package manager, `pip`, through the following terminal command:

```
(venv)$ pip install -r requirements.txt
```

Then it is necessary to set up the Material Project "API keys" to enable the code to query the database. To obtain the API key, follow the instructions provided in the documentation, which can be found here. After obtaining the API key, replace `<MATERIAL_PROJECT_KEY>` with your actual API key and run the script as shown below:

```
(venv)$ python tools/mpinterface_setup.py -k <MATERIAL_PROJECT_KEY>
```

Finally, to utilize the Jupyter Notebooks, it is necessary to run the Jupyter Lab server using the following command:

```
(venv)$ jupyter lab
```

## 1.2 Folder Structure

The repository is structured as follows:

```
.
├── assets
├── data
│   └── ffield.reax.optimized.ff
├── tools
│   └── ...
├── ...
├── CREDITS.md
├── LICENSE
├── README.md
├── requirements.txt
├── JNB1-Initial_configrations.ipynb
├── JNB2-Simulations.ipynb
├── JNB3-Build_traingset.ipynb
├── preJNB4-ReaxFF_optimization.ipynb
└── JNB4-ReaxFF_optimization.ipynb
```

In the folder tree, we have excluded certain auxiliary files and folders that are not essential for understanding the repository, and below we explain in detail the content of each folder/file:

- `assets`: This folder contains any additional assets, such as images or documentation; related to the repository.

- `data`: This folder contains the data files resulting from this work.

  - `ffield.reax.optimized.ff`: This file is the optimized ReaxFF resulting from using these Jupyter notebooks, as explained in the main text, and as reported in Section 5.

- `tools`: This directory contains a collection of Python modules and scripts.

- `CREDITS.md`: This file acknowledges and credits each author who contributed to the repository.

- `LICENSE`: This file contains the license information for the repository (CC BY 4.0). It specifies the terms and conditions under which the repository's contents are distributed and used.

- `README.md`: This is the "readme" file (repository overview and instructions).

- `requirements.txt`: This file lists the required Python packages and their versions (see Subsection [1.1](#)).

- `JNB1-Initial_configurations.ipynb`: Jupyter Notebook where the protocol is initialized by querying the *Materials Project* database, downloading the unit crystals, and producing all the initial configurations for the DFT simulations using the *pymatgen* library.

- `JNB2-Simulations.ipynb`: This notebook performs the simulations using BAND and DFTB codes available in the Amsterdam Modeling Suite. The simulations are performed in parallel using the *PLAMS* library and *SLURM* scheduler.

- `JNB3-Build_trainingset.ipynb`: Here, the quantities needed for the database are extracted and tuned to favor accuracy on the energy.

- `preJNB4-ReaxFF_optimization.ipynb`: This is an auxiliary Notebook where the old ReaxFF is converted into a Python object, and it is possible to select the subset of coefficients related to specific interactions to change during the optimization (e.g., bond, van der Waals, angular, etc.).

- `JNB4-ReaxFF_optimization.ipynb`: This notebook takes the database and the ReaxFF Python object to perform a multi-objective optimization and find the new ReaxFF potential that minimizes the Sum of Squared Errors (SSE).

## 2 Enhancing ReaxFF database (Database Repository)

In addition to the protocol, we are sharing the complete database used to reparameterize ReaxFF. This allows for testing the optimization without rerunning time-consuming simulations and enables future expansion and improvement by the community. We stored the data using the ASE SQLite3 database, which was chosen for its widespread use in the computational community of European institutions. The use of this type database provides fast and flexible querying and browsing capabilities, as described in Subsection 2.2. The database contains information on initial configurations, single-point calculations, pre-optimizations, and DFT optimizations. However, only a subset of the database is used in the training set for the ReaxFF optimization, as explained in the main text and briefly summarized in Table S1. For each entry in the database, we store various information regarding the simulation and properties obtained from the DFT calculations. These details are listed in Table S2.

**Table S1.** Training set composition, with quantities calculated using DFT: Energy (E), Forces (F), and Charges (Q). The numbers in parentheses indicate the successful cases for self-consistent field (SCF) calculations that may not always converge.

| Type | Unit cell | | | Quantities | | |
|---|---|---|---|---|---|---|
| | $Fm\bar{3}m$ | $P6_3mc$ | $Pm\bar{3}m$ | E | F | Q |
| Supercells | 6 | 6 | 6 | ✓ | ✓ | ✓ |
| Vacancies | 10 | 24 | 10 | ✓ | | ✓ |
| Strain | 39 | 39 | 39 | ✓ | | ✓ |
| Substitution | 5 | 5 | 5 | ✓ | | ✓ |
| Interstitial | 5 | 5 | 5 | ✓ | ✓ | ✓ |
| Slabs | 12(10) | 12(8) | 12(11) | ✓ | ✓ | ✓ |
| Bulk 300 K | 10 | 10 | 10 | ✓ | | |
| Bulk 500 K | 10 | 10 | 10 | ✓ | | |
| Amorphous | 10 | 10 | 10 | ✓ | | |
| Total | 107(105) | 121(117) | 107(106) | | | |

The repository is available both in Zenodo at the permanent link `https://doi.org/10.5281/zenodo.7959121`, and on version control repositories host GitHub `https://github.com/paolodeangelis/Enhancing_ReaxFF_DFT_database`, Figure S3.

**Figure S3.** Screenshot of the database hosted on GitHub (a) and Zenodo (b).

**Table S2.** Datailed list of properties and information store in each database entry

| name | description | type | unit |
|---|---|---|---|
| `id` | Uniqe row ID | integer | — |
| `ctime` | Creation time of the data (for simulation, it coincides with the runtime) | float | yr |
| `formula` | Chemical formula of the system | string | — |
| `pbc` | Periodic boundary conditions | boolean | — |
| `user` | Username or full name of the user who created the data | string | — |
| `calculator` | Name of the ASE-calculator and engine used to calculate the system (e.g `ams/band` means AMS calculator with BAND engine) | string | — |
| `energy` | Total energy of the system from the calculation | float | eV |
| `natoms` | Number of atoms | integer | — |
| `fmax` | Maximum force | float | eV Å$^{-1}$ |
| `smax` | Maximum stress | float | eV/Å$^3$ |
| `charge` | Net charge in unit cell | float | $|e|$ |
| `mass` | Sum of atomic masses in unit cell | float | au |
| `magmom` | Magnetic moment | float | $\mu_B$ |
| `unique_id` | Random (unique) ID | integer | — |
| `volume` | Volume of unit cell | float | Å$^3$ |
| `functional` | Exchange-and-correlation functional | string | — |
| `fermi_energy` | Fermi Energy (N.B. is not the Fermi Level), which indicates the energy of non-interacting fermions in the system (Fermi gas) at $0\,\mathrm{K}$ | float | eV |
| `homo_energy` | Highest Occupied Molecular Orbital energy | float | eV |
| `lumo_energy` | Lowest Unoccupied Molecular Orbital energy | float | eV |
| `band_gap` | Band gap energy (for LiF is the HOMO energy - LUMO energy) | float | eV |
| `runtime` | Simulation start date whit format `%b%d-%Y %H:%M:%S` | string | — |
| `elapsed` | Elapsed Time | float | s |
| `name` | System Name | string | — |
| `sim_name` | Simulation Name | string | — |
| `subset_name` | Name of the subset of configuration | string | — |
| `run_script` | Full AMS scrip for running the simulation | string | — |
| `input_script` | Full input scrip for running the simulation | string | — |
| `success` | Simulation end status | boolean | — |
| `used_in` | Indicate in which set (training-set or test-set) the data was used | string | — |
| `task` | Simulation task type | string | — |
| `space_group` | Full International Space Group Symbol. The notation is a LaTeX-like string, with screw axes being represented by an underscore | string | — |
| `data` | Additional data calculate: Density of states(DOS) and Hystroy (e.g. energy, force evolution during the simulation) | dictionary | — |

## 2.1 Installation

To browse and query the database, only Python and the ASE and Jupyter Lab libraries are required (see Subsubsection 2.1.1). With this minimal setup, it is possible to perform additional simulations using the preferred DFT engine. Using ASE, you can wrap the simulations and add them to the database. For an example, you can refer to the `notebooks\running_simulation.ipynb` Jupyter Notebook, which demonstrates the usage of the BAND plane-wave DFT code. However, if you wish to rerun the simulations stored in the database, the commercial code Amsterdam Modeling Suite (AMS) is required (refer to Subsubsection 2.1.2).

### 2.1.1 Minimum Requirements
- Python 3.8 or above
- Atomic Simulation Environment (ASE) library
- Jupyter Lab

### 2.1.2 Requirements for Re-running or Performing New Simulations
- SCM (Software for Chemistry & Materials) Amsterdam Modeling Suite
- PLAMS (Python Library for Automating Molecular Simulation) library

### 2.1.3 Python project setup
Similar to the protocol repository, the project setup requires the installation of additional Python packages such as ASE, PLAMS, etc. These packages can be installed using the Python package manager, `pip`, with the following terminal command:

```
(venv)$ pip install -r requirements.txt
```

Then it is possible to utilize the Jupyter Notebooks by starting the Jupyter Lab server using the following command:

```
(venv)$ jupyter lab
```

## 2.2 Interacting with the Database

There are three methods available for interacting with the database: using the ASE db command line, the web interface, and the ASE Python interface.

### 2.2.1 ASE db Command-line
To interact with the database through the ASE db command line, follow these steps:

1. Open a terminal and navigate to the directory where the `LiF.db` file is located.
2. Execute the following command to initiate the ASE db terminal:

   ```
   (venv)$  ase db LiF.db
   ```

3. Now, one can utilize the available commands within the terminal to query and manipulate the database.

**Figure S4.** Example of interacting with the database via the terminal.

#### 2.2.2 ASE Python Interface

To interact with the database using the ASE Python interface, you can utilize the following code example:

```python
from ase.db import connect

# Connect to the database
db = connect("LiF.db")

# Query the database
results = db.select('success=True')

# Iterate over the results
for row in results:
    print(f"ID: {row.id}, Energy: {row.energy}")
```

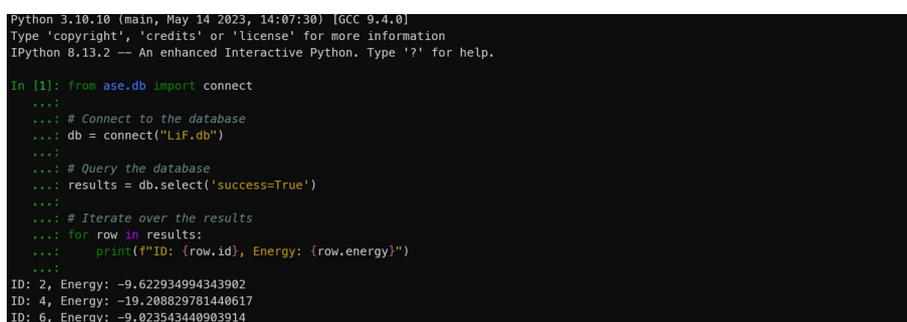

**Figure S5.** Example of interacting with the database via Python, utilizing the iPython terminal interface.

For a more detailed example, refer to the notebook `notebooks\browsing_db.ipynb`. To learn how to perform a simulation, check the notebook `notebooks\running_simulation.ipynb`.

#### 2.2.3 Web Interface

To interact with the database using the web interface, follow these steps:

1. Open a terminal and navigate to the directory where the `LiF.db` file is located.
2. Execute the following command to initiate the ASE db terminal:

    ```
    (venv)$ ase db -w LiF.db
    ```

3. Access the web browser and connect to the local server at `http://127.0.0.1:5000`.
4. Upon accessing the interface, Figure S6.a will be displayed, showcasing all the entries in the database.
5. To query the database and filter the entries based on specific criteria, users can input their search parameters into the designated fields (e.g. `success=True`).
6. Once the query is executed, Figure S6.b will be displayed, presenting the selected entries that match the search criteria.
7. To access all the information of an individual entry, users can click on a specific entry to view a detailed visualization. Figure S7 demonstrates an example of a single entry displayed through the web interface.

**Figure S6.** Screenshot showing the web interface of the database, displaying all the entries (a), and filtering the entries after querying the database (b).

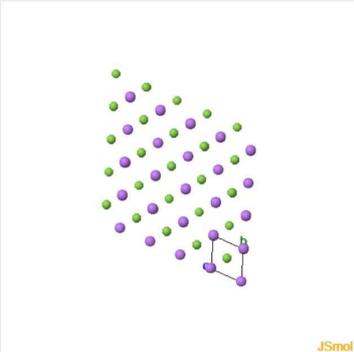

**Figure S7.** Screenshot displaying a entrie from the database visualized using the web interface.

## 2.3 Folder Structure
The repository is structured as follows:

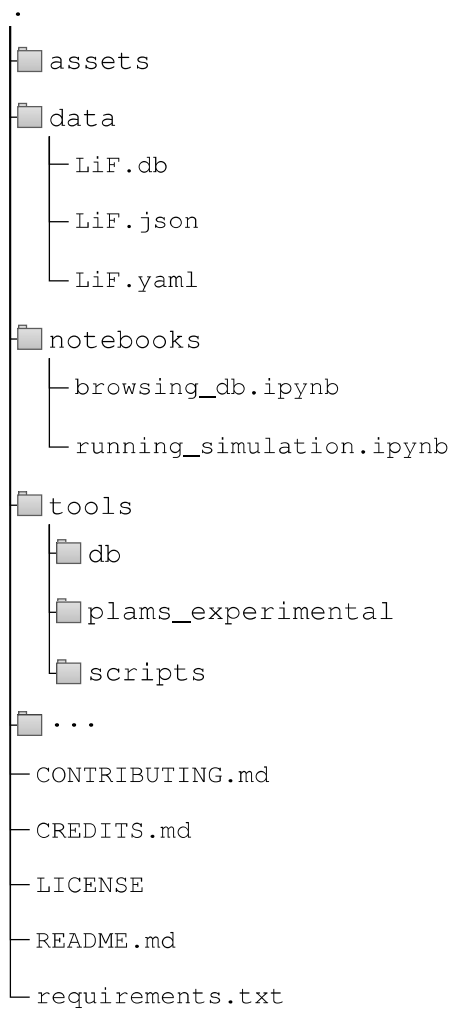

```
.
├── assets
├── data
│   ├── LiF.db
│   ├── LiF.json
│   └── LiF.yaml
├── notebooks
│   ├── browsing_db.ipynb
│   └── running_simulation.ipynb
├── tools
│   ├── db
│   ├── plams_experimental
│   └── scripts
├── ...
├── CONTRIBUTING.md
├── CREDITS.md
├── LICENSE
├── README.md
└── requirements.txt
```

- `assets`: This folder contains any additional assets, such as images or documentation, related to the repository.

- `data`: This folder contains the data files used in the repository.

  - `LiF.db`: This file is the SQLite database file that includes the DFT data used for the ReaxFF force field. Specifically, it contains data related to the inorganic compound LiF.

  - `LiF.json`: This file provides the database metadata in a human-readable format JSON.

  - `LiF.yaml`: This file also contains the database metadata in a more human-readable format YAML.

- `notebooks`: This folder contains Jupyter notebooks that provide demonstrations and examples of how to use and analyze the database.

  - `browsing_db.ipynb`: This notebook demonstrates how to handle, select, read, and understand the data points in the `LiF.db` database using the ASE database Python interface. It serves as a guide for exploring and navigating the database effectively.

- – `running_simulation.ipynb`: In this notebook, you will find an example of how to get a data point from the `LiF.db` database and use it to perform a new simulation. The notebook showcases how to utilize either the PLAMS library or the AMSCalculator and ASE Python library to conduct simulations based on the retrieved data and then store it as a new data point in the `LiF.db` database. It provides step-by-step instructions and code snippets for a seamless simulation workflow.

- `tools`: This directory contains a collection of Python modules and scripts that are useful for reading, analyzing, and re-running simulations stored in the database. These tools are indispensable for ensuring that this repository adheres to the principles of **I**nteroperability and **R**eusability, as outlined by the FAIR principles.

  - `db`: This Python module provides functionalities for handling, reading, and storing data into the database.
  
  - `plasm_experimental`: This Python module includes the necessary components for using the `AMSCalculator` with PLASM and the SCM software package, utilizing the ASE API. It facilitates running simulations, and performing calculations.
  
  - `scripts`: This directory contains additional scripts for advanced usage scenarios of this repository.

- `CONTRIBUTING.md`: This file provides guidelines and instructions for contributing to the repository. It outlines the contribution process, coding conventions, and other relevant information for potential contributors.

- `CREDITS.md`: This file acknowledges and credits each author who contributed to the repository.

- `LICENSE`: This file contains the license information for the repository (CC BY 4.0). It specifies the terms and conditions under which the repository's contents are distributed and used.

- `README.md`: This is the "readme" file (repository overview and instructions).

- `requirements.txt`: This file lists the required Python packages and their versions. (see Subsection 2.1)

## 3 DFT simulations

In this section, we present the results of our DFT simulations, highlighting the energy obtained for various configurations.

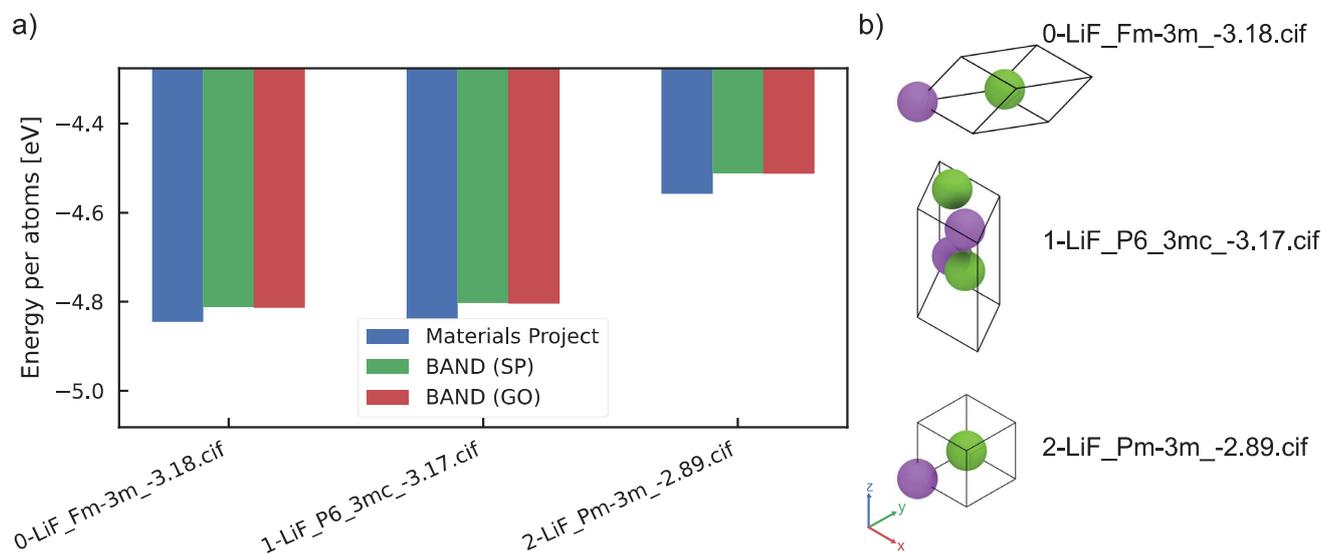

**Figure S8.** Comparison of energy per atom obtained from DFT simulations (SP: Single Point, GO: Geometry Optimization) with those stored on Material Project (a), for each primitive unit cell of LiF (b).

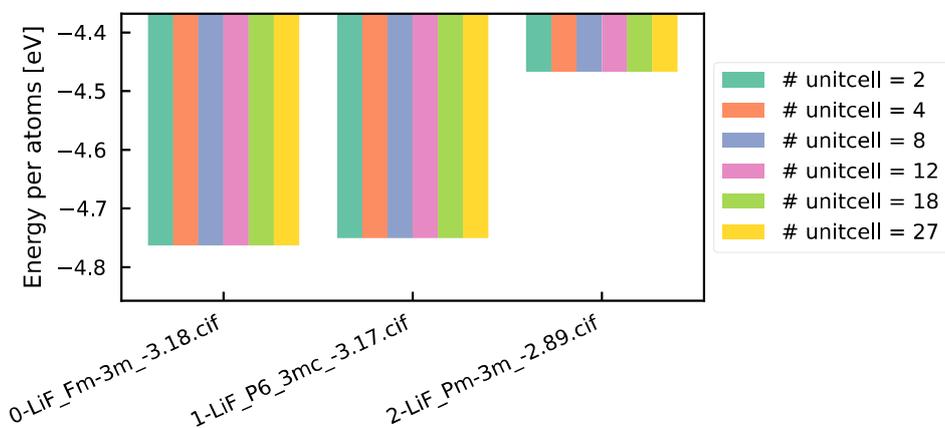

**Figure S9.** Energy per atom obtained from DFT simulations for each crystal in the LiF supercell.

**Figure S10.** Energy per atom curves obtained from strained configurations, specifically with 1D strain $\varepsilon_{11}$, shear strain $\varepsilon_{12} = \varepsilon_{21}$, and homogeneous expansion/compression $\varepsilon_{11} = \varepsilon_{22} = \varepsilon_{33}$.

**Figure S11.** Energy per atom obtained from DFT simulations for each system where a vacancy defect was added.

**Figure S12.** Energy per atom obtained from DFT simulations for each system where a substitution defect was added.

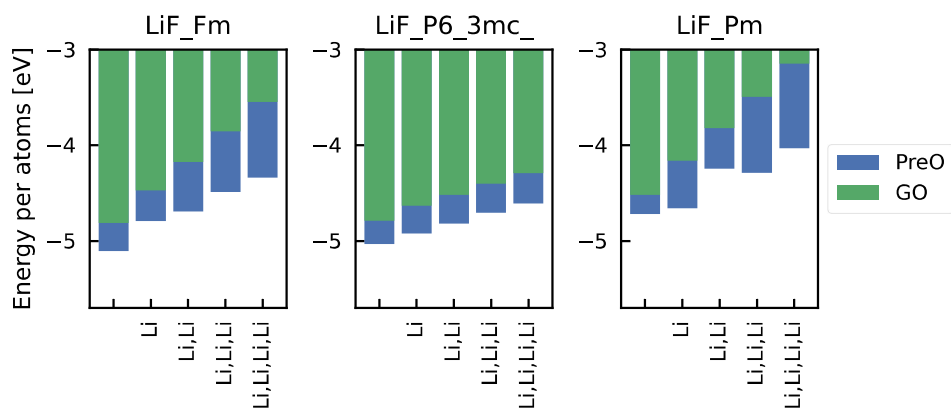

**Figure S13.** Energy per atom obtained from DFT simulations for each system with an added interstitial defect. Due to the significant crystal deformation introduced, the optimization process was conducted in two steps: Pre-optimization (blue) and full geometry optimization (green).

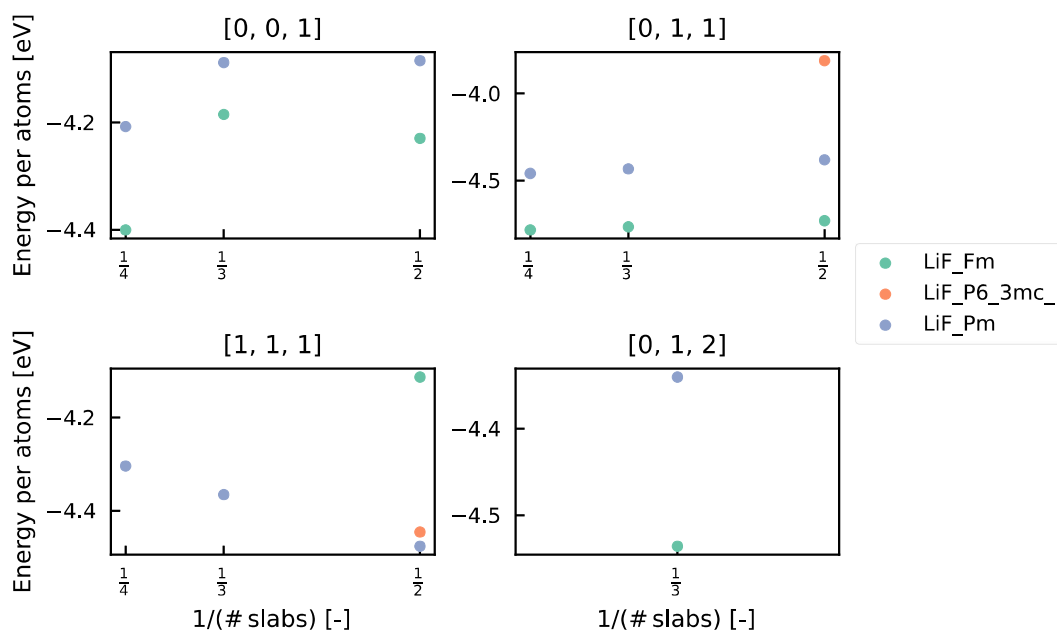

**Figure S14.** Comparing energy per atom for different crystal plates ($[0,0,1]$, $[0,1,1]$, $[1,1,1]$, $[0,1,2]$) with varying slab thickness. The plot illustrates how the energy per atom changes as a function of the inverse of the number of slabs, emphasizing the impact of surface energy.

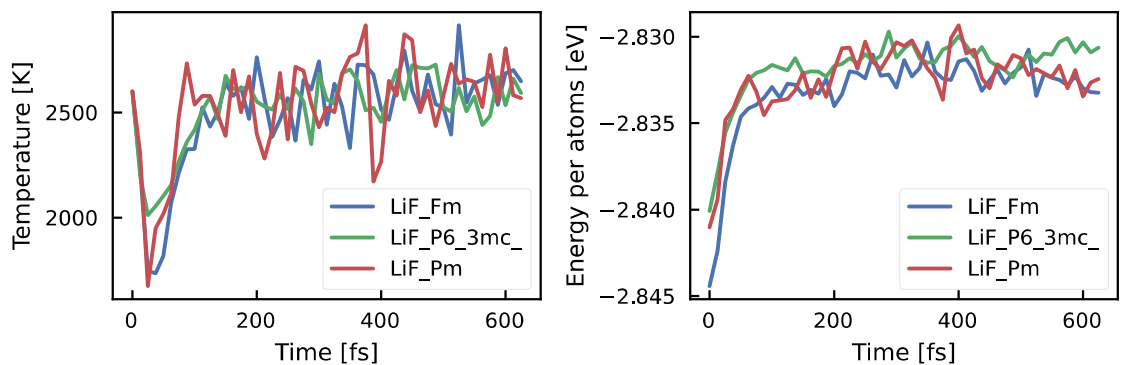

**Figure S15.** Evolution of temperature and energy per atom in *ab initio* NVT-MD simulation at 2500 K with Non-Scc-GFN1xTB DFTB model[6] for sampling amorphous LiF systems.

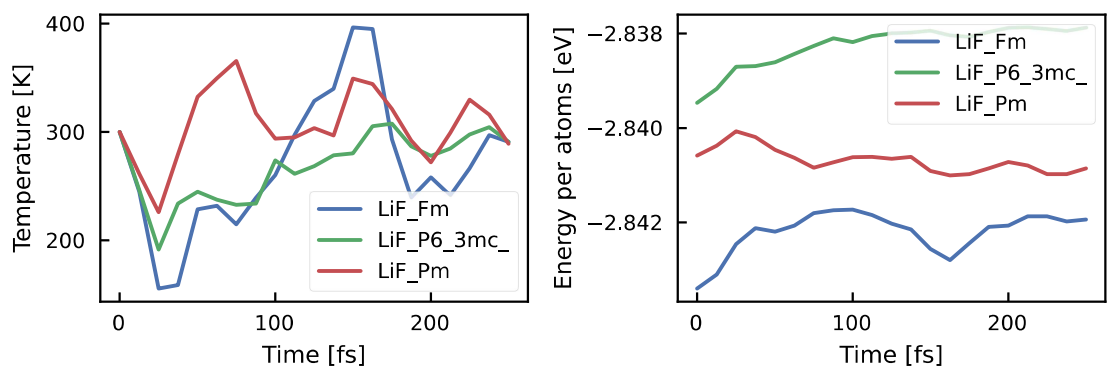

**Figure S16.** Evolution of temperature and energy per atom in *ab initio* NVT-MD simulation at 300 K with Non-Scc-GFN1xTB DFTB model[6] for sampling solid LiF systems.

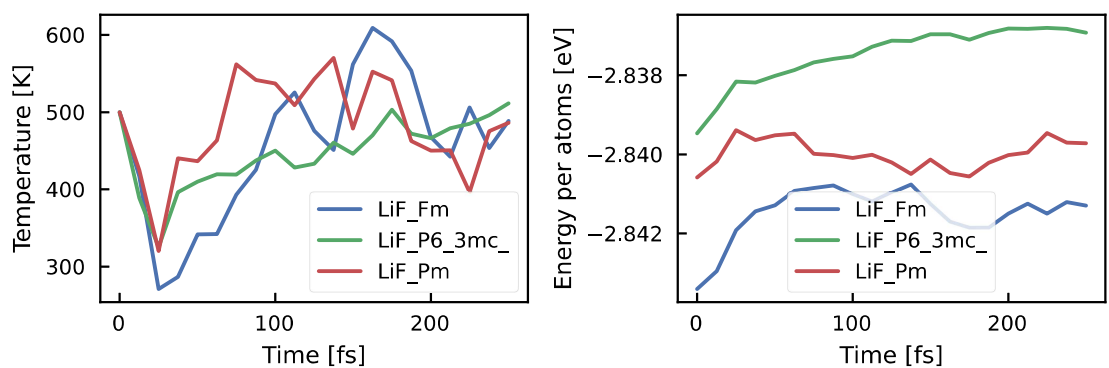

**Figure S17.** Evolution of temperature and energy per atom in *ab initio* NVT-MD simulation at 500 K with Non-Scc-GFN1xTB DFTB model[6] for sampling solid LiF systems.

## 3.1 DFT vs. ReaxFF mechanical deformation

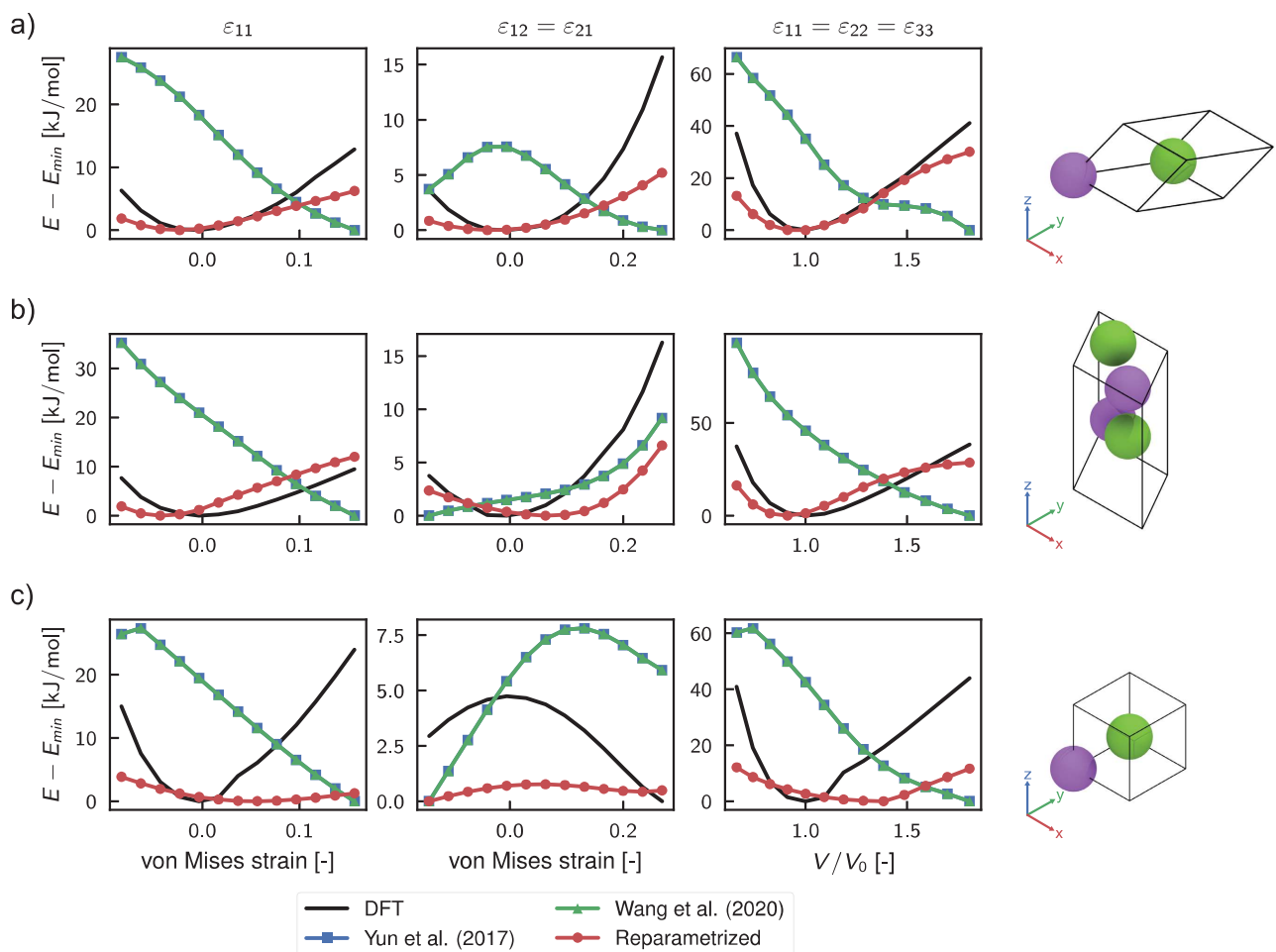

**Figure S18.** Comparing the mechanical response of LiF: ReaxFF and DFT Predictions. The black line represents DFT results used for training, while the blue, green, and red lines depict the energy predictions by ReaxFF models from Yun et al.[7], Wang et al.[8], and our proposed new reparameterization, respectively. We examine the energy variation with respect to the equilibrium crystal for three types of deformation: tensile strain deformation $\varepsilon_{11}$ (left), shear strain deformation $\varepsilon_{12} = \varepsilon_{21}$ (center), and homogeneous lattice length change $V/V_0$ (where $V_0$ is the equilibrium volume). This study was performed on all stable and metastable crystals with space groups $Fm\bar{3}m$ (a), $Pm\bar{3}m$ (b), and $P6_3mc$ (c) from the Material Project database[9].

## 4 Diffusion MSD analysis

Presented here is a plot illustrating the Mean Square Displacement (MSD) analysis conducted to investigate Li diffusion using ReaxFF simulatios.

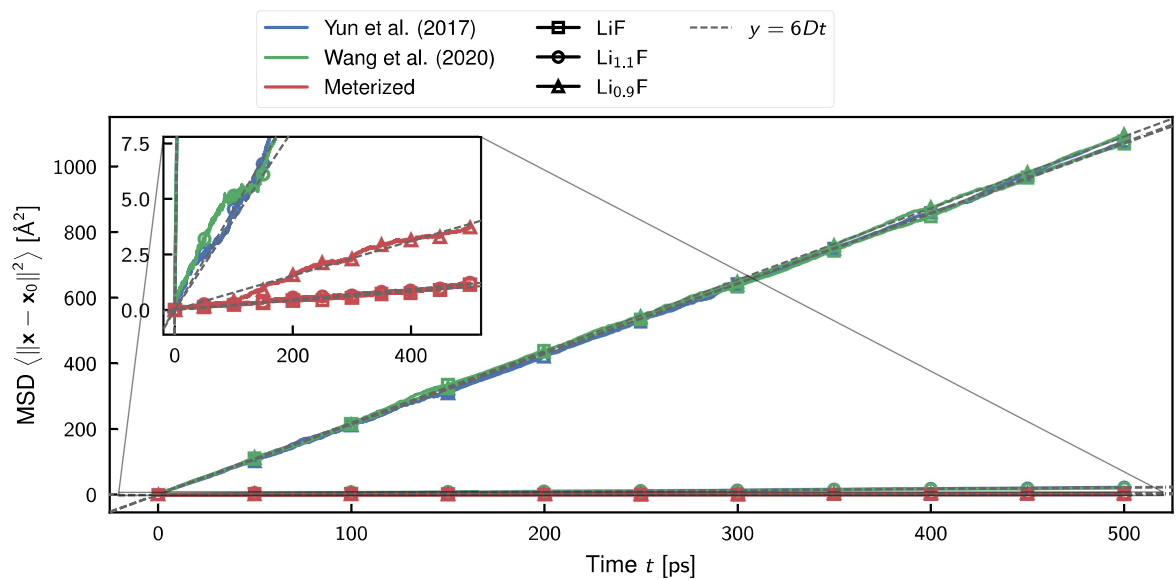

**Figure S19.** Evolution of Mean Square Displacement (MSD) during NVT-MD ReaxFF simulations of Pure Bulk LiF (LiF) and point defects: vacancy ($Li_{0.9}F$) and interstitial ($Li_{1.1}F$) at 300 K. The markers (square, circle, and triangular) represent the corresponding systems in the plot. The simulations were performed using ReaxFF parameterizations by Yun et al.[7] (blue), Wang et al.[8] (green), and our proposed new parameterization (red). The dashed line indicates the linear model ($MSD = 6D \cdot t$) employed to compute diffusivity in each simulation.

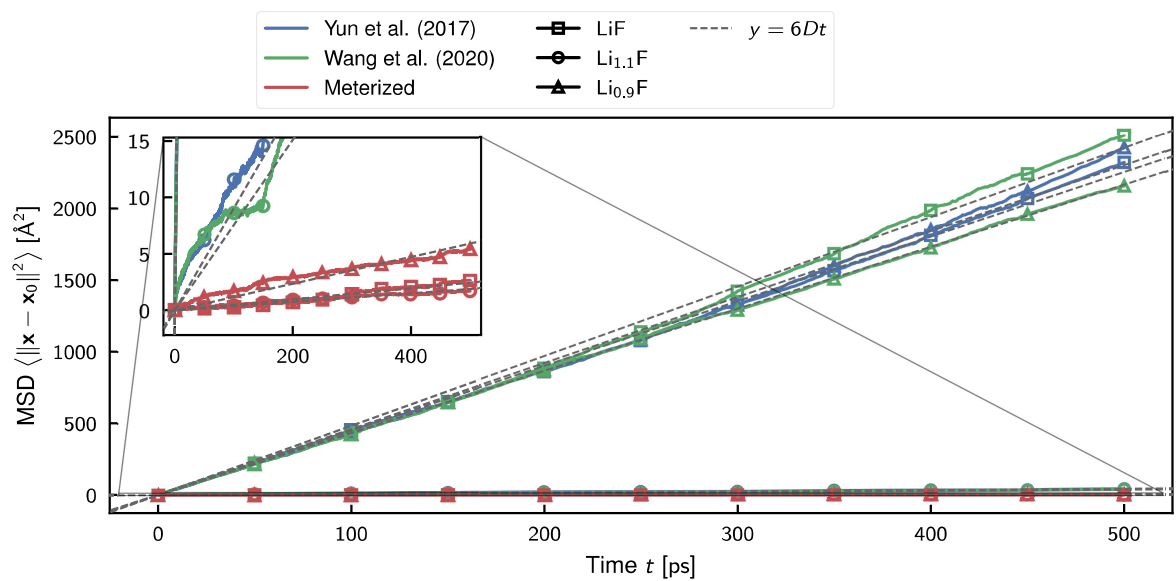

**Figure S20.** Evolution of Mean Square Displacement (MSD) during NVT-MD ReaxFF simulations of Pure Bulk LiF (LiF) and point defects: vacancy ($Li_{0.9}F$) and interstitial ($Li_{1.1}F$) at 400 K. The markers (square, circle, and triangular) represent the corresponding systems in the plot. The simulations were performed using ReaxFF parameterizations by Yun et al.[7] (blue), Wang et al.[8] (green), and our proposed new parameterization (red). The dashed line indicates the linear model ($MSD = 6D \cdot t$) employed to compute diffusivity in each simulation.

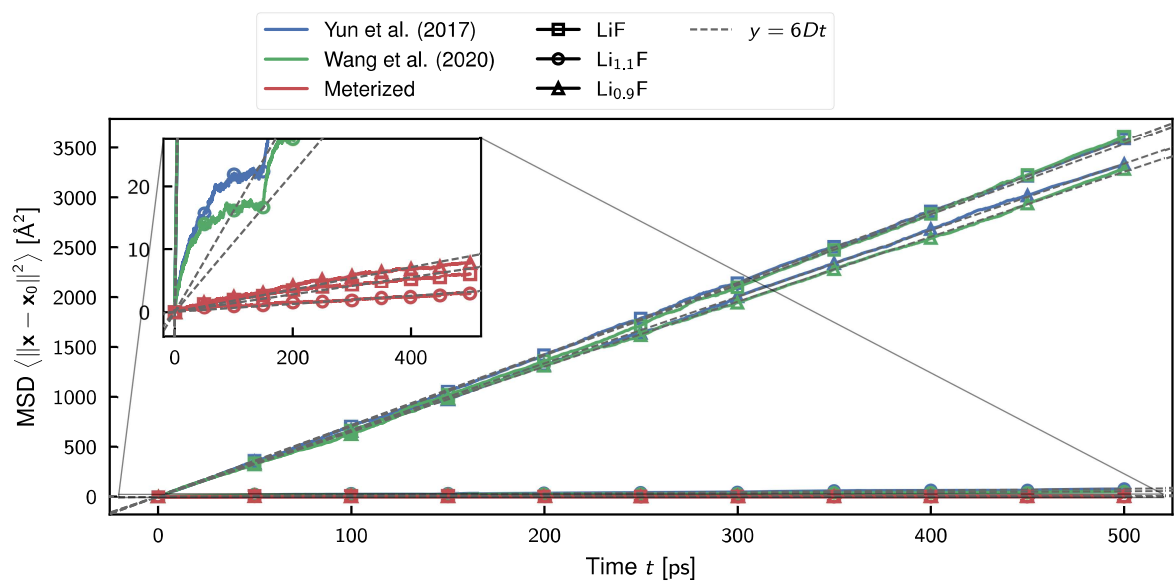

**Figure S21.** Evolution of Mean Square Displacement (MSD) during NVT-MD ReaxFF simulations of Pure Bulk LiF (LiF) and point defects: vacancy ($Li_{0.9}F$) and interstitial ($Li_{1.1}F$) at 500 K. The markers (square, circle, and triangular) represent the corresponding systems in the plot. The simulations were performed using ReaxFF parameterizations by Yun et al.[7] (blue), Wang et al.[8] (green), and our proposed new parameterization (red). The dashed line indicates the linear model ($MSD = 6D \cdot t$) employed to compute diffusivity in each simulation.

## 5 Reparameterized ReaxFF

```
 1  ! ReaxFF force field C/H/O/Si/Li/F by P. De Angelis (reparameterized from ReaxFF  by Yun,
        Kang-Seop, et al. (2017).).
 2   39       ! Number of general parameters
 3     50.0000 !Overcoordination parameter
 4      9.5469 !Overcoordination parameter
 5     26.5405 !Valency angle conjugation parameter
 6      1.7224 !Triple bond stabilisation parameter
 7      6.8702 !Triple bond stabilisation parameter
 8     60.4850 !C2-correction
 9      1.0588 !Undercoordination parameter
10      4.6000 !Triple bond stabilisation parameter
11     12.1176 !Undercoordination parameter
12     13.3056 !Undercoordination parameter
13    -70.5044 !Triple bond stabilization energy
14      0.0000 !Lower Taper-radius
15     10.0000 !Upper Taper-radius
16      2.8793 !Not used
17     33.8667 !Valency undercoordination
18      6.0891 !Valency angle/lone pair parameter
19      1.0563 !Valency angle
20      2.0384 !Valency angle parameter
21      6.1431 !Not used
22      6.9290 !Double bond/angle parameter
23      0.3989 !Double bond/angle parameter: overcoord
24      3.9954 !Double bond/angle parameter: overcoord
25     -2.4837 !Not used
26      5.7796 !Torsion/BO parameter
27     10.0000 !Torsion overcoordination
28      1.9487 !Torsion overcoordination
29     -1.2327 !Conjugation 0 (not used)
30      2.1645 !Conjugation
31      1.5591 !vdWaals shielding
32      0.0010 !Cutoff for bond order (*100)
33      2.1365 !Valency angle conjugation parameter
34      0.6991 !Overcoordination parameter
35     50.0000 !Overcoordination parameter
36      1.8512 !Valency/lone pair parameter
37      0.5000 !Not used
38     20.0000 !Not used
39      5.0000 !Molecular energy (not used)
40      0.0000 !Molecular energy (not used)
41      2.6962 !Valency angle conjugation parameter
42   6    ! Nr of atoms; cov.r; valency; a.m.; Rvdw; Evdw; gammaEEM; cov.r2; #el.
43        ! alfa; gammavdW; valency13; Eunder; Eover; chiEEm; etaEEM; n.u1
44        ! cov.r3; Elp; Heat inc.; 13BO1; 13BO2; 13BO3; n.u2; n.u3
45        ! ov/un; vval1;  vval2;  vval3;  vval4;  n.u5; n.u6; n.u7
46  C   1.3825   4.0000  12.0000   1.9133   0.1853   0.9000   1.1359   4.0000
47      9.7602   2.1346   4.0000  33.2433  79.5548   5.8678   7.0000   0.0000
48      1.2104   0.0000 199.0303   8.6991  34.7289  13.3894   0.8563   0.0000
49     -2.8983   2.5000   1.0564   4.0000   2.9663   0.0000   0.0000   0.0000
50  H   0.7853   1.0000   1.0080   1.5904   0.0419   1.0206  -0.1000   1.0000
51      9.3557   5.0518   1.0000   0.0000 121.1250   5.3200   7.4366   1.0000
52     -0.1000   0.0000  62.4879   1.9771   3.3517   0.7571   1.0698   0.0000
53    -15.7683   2.1488   1.0338   1.0000   2.8793   0.0000   0.0000   0.0000
54  O   1.2450   2.0000  15.9990   2.3890   0.1000   1.0898   1.0548   6.0000
55      9.7300  13.8449   4.0000  37.5000 116.0768   8.5000   8.3122   2.0000
56      0.9049   0.4056  59.0626   3.5027   0.7640   0.0021   0.9745   0.0000
57     -3.5500   2.9000   1.0493   4.0000   2.9225   1.3000   0.2000  13.0000
```

```
 58   Si    2.2902    4.0000   28.0600    1.8354    0.2110    0.5947    1.2962    4.0000
 59        11.1336    3.1831    4.0000   21.7115  139.9309    4.2033    5.5558    0.0000
 60        -1.0000    0.0000  104.0000    9.0751   23.8188    0.8381    0.8563    0.0000
 61        -4.1684    2.0754    1.0338    4.0000    2.5791    1.4000    0.2000   13.0000
 62   Li    1.9205    1.0000    6.9410    1.8896    0.0905    0.4668   -0.1000    1.0000
 63         9.9084    1.0896    1.0000    0.0000    0.0000   -6.4188   15.0000    0.0000
 64        -1.0000    0.0000   37.5000    5.4409    6.9107    0.1973    0.8563    0.0000
 65       -25.0000    2.2989    1.0338    1.0000    2.8103    1.3000    0.2000   13.0000
 66   F     1.2705    1.0000   18.9984    1.4100    0.0442    0.0135   -0.1000    7.0000
 67        11.6107    5.4481    4.0000   62.1473    0.1252   15.4329   17.3228    0.0000
 68        -1.0000   35.0000    1.5000    6.9821    4.1799    1.0561    0.0000    0.0000
 69        -2.2869    2.7340    1.0493    4.0000    3.0013    0.0000    0.0000    0.0000
 70   21      ! Nr of bonds; Edis1; Edis2; Edis3; pbe1;   pbo5; 13corr; pbo6; kov
 71           ! pbe2; pbo3;   pbo4; n.u1; pbo1; pbo2; ovcorr; n.u2
 72    1  1 156.5953 100.0397   80.0000   -0.8157   -0.4591    1.0000   37.7369    0.4235
 73           0.4527   -0.1000    9.2605    1.0000   -0.0750    6.8316    1.0000    0.0000
 74    1  2 170.2316    0.0000    0.0000   -0.5931    0.0000    1.0000    6.0000    0.7140
 75           5.2267    1.0000    0.0000    1.0000   -0.0500    6.8315    0.0000    0.0000
 76    1  4  94.5912   50.1197    0.0000   -0.5712   -0.5558    1.0000   17.2117    0.0308
 77           2.3951   -1.1892    8.6403    1.0000   -0.1028    5.4278    1.0000    0.0000
 78    2  2 156.0973    0.0000    0.0000   -0.1377    0.0000    1.0000    6.0000    0.8240
 79           2.9907    1.0000    0.0000    1.0000   -0.0593    4.8358    0.0000    0.0000
 80    1  3 224.8293   31.8847   89.5456   -1.4925   -0.0850    1.0000   13.4838    1.1008
 81           0.7387   -0.7228    5.1953    1.0000   -0.1175    6.4319    0.0000    0.0000
 82    3  3 142.2858  145.0000   50.8293    0.2506   -0.1000    1.0000   29.7503    0.6051
 83           0.3451   -0.1055    9.0000    1.0000   -0.1225    5.5000    1.0000    0.0000
 84    2  3 160.0000    0.0000    0.0000   -0.5725    0.0000    1.0000    6.0000    0.5626
 85           1.1150    1.0000    0.0000    0.0000   -0.0920    4.2790    0.0000    0.0000
 86    2  4 101.1840    0.0000    0.0000   -0.1751    0.0000    1.0000    6.0000    1.1044
 87           7.3549    1.0000    0.0000    1.0000   -0.0450    7.9080    0.0000    0.0000
 88    3  4 274.8339    5.0000    0.0000   -0.5884   -0.3000    1.0000   36.0000    0.2131
 89           9.9772   -0.2572   28.8153    1.0000   -0.1130    8.4790    6.0658    0.0000
 90    4  4  61.1127   85.8146   30.0000   -0.8197   -0.3000    1.0000   16.0000    0.1386
 91           0.1307   -0.8055    7.1248    1.0000   -0.0674    8.2374    0.0000    0.0000
 92    1  5  10.0540    0.0000    0.0000    0.3005   -0.3000    0.0000    6.0000    0.2953
 93           0.2679   -0.2534   12.0019    1.0000   -0.1143    7.5211    0.0000    0.0000
 94    2  5  63.4649    0.0000    0.0000    0.0294    0.0000    0.0000    6.0000    0.4868
 95           0.3090    0.0000   12.0000    1.0000   -0.0800    5.1033    0.0000    0.0000
 96    5  3  78.3666   -0.0200    0.0000   -1.0000    0.3000    0.0000    6.0000    0.3228
 97           0.2022   -0.2500   11.9965    1.0000   -0.1276    7.8656    0.0000    0.0000
 98    5  4  23.1963    0.0000    0.0000    1.0000    0.3000    0.0000   26.0000    0.5185
 99           0.0812    0.0000   12.0000    1.0000   -0.1142    6.0525    0.0000    0.0000
100    5  5  16.7443    0.0000    0.0000    1.0420    0.3000    0.0000   26.0659    0.5820
101           1.6137   -0.0805   12.0000    1.0000   -0.1646    4.7897    0.0000    0.0000
102    1  6 166.8800    0.0000    0.0000    0.4753   -0.5000    1.0000   35.0000    1.3151
103           3.1303   -0.2500   15.0000    1.0000   -0.9475    7.1188    1.0000    0.0000
104    6  2 260.6892    0.0000    0.0000   -0.6690    0.0000    1.0000    6.0000    2.3211
105           8.8937    1.0000    0.0000    1.0000   -0.3966    9.2031    0.0000    0.0000
106    6  3  99.9065    0.0000    0.0000   -1.0000    0.0000    1.0000    6.0000    0.7194
107           0.5928    1.0000    0.0000    1.0000   -0.1318    8.4278    0.0000    0.0000
108    6  6  65.8563    0.0000    0.0000   -0.0630   -0.5267    1.0116   30.2616    0.2245
109          -0.2306   -0.2161   15.3928    1.0000   -0.1007    8.6628    0.0597    0.0000
110    6  5  24.7086    0.0000    0.0000   -1.3775   -0.0961    0.4644   45.9272    0.6224
111           6.9172   -0.0851   14.8066    1.0000   -0.0752    9.0692    0.7043    0.0000
112    6  4 284.8610    0.0000    0.0000   -0.8680   -0.5000    1.0000   35.0000    1.4117
113           3.5449   -0.2500   15.0000    1.0000   -0.1198    6.0380    1.0000    0.0000
114   15      ! Nr of off-diagonal terms; Ediss; Rvdw; alfa; cov.r; cov.r2; cov.r3
115    1  2    0.1219    1.4000    9.8442    1.1203   -1.0000   -1.0000
116    1  3    0.1893    1.7076   10.2970    1.3608    1.0384    1.0646
117    1  4    0.5876    1.3349   13.4198    1.4988    1.6946   -1.0000
```

```
118    2  3   0.0283   1.2885  10.9190   0.9215  -1.0000  -1.0000
119    2  4   0.1035   1.3327  11.5963   1.3977  -1.0000  -1.0000
120    3  4   0.1836   1.9157  10.9070   1.7073   1.2375  -1.0000
121    1  5   0.0610   1.5665  11.4404   1.2079  -1.0000  -1.0000
122    2  5   0.2966   1.2550  10.2920   1.1989  -1.0000  -1.0000
123    5  3   0.0790   2.2000   9.0491   1.8165  -1.0000   1.0000
124    5  4   0.1271   2.0090  11.5659   1.7156   1.0000   1.0000
125    1  6   0.0749   1.7702  12.0430   1.6315  -1.0000  -1.0000
126    6  2   0.0702   1.1145  12.1833   1.0722  -1.0000  -1.0000
127    6  3   0.1293   1.3773  11.4683   1.3410  -1.0000  -1.0000
128    6  5   0.0495   1.2573  15.6983   0.9415  -1.0000  -1.0000
129    6  4   0.0996   1.5863  13.3401   1.5077  -1.0000  -1.0000
130   52      ! Nr of angles; Theta0; ka; kb; pconj; pv2; kpenal; pv3
131    1  1  1  67.2326  22.0695  1.6286  0.0000  1.7959  15.4141  1.8089
132    1  1  2  65.2527  14.3185  6.2977  0.0000  0.5645   0.0000  1.1530
133    2  1  2  70.0840  25.3540  3.4508  0.0000  0.0050   0.0000  3.0000
134    1  2  2   0.0000   0.0000  6.0000  0.0000  0.0000   0.0000  1.0400
135    1  2  1   0.0000   3.4110  7.7350  0.0000  0.0000   0.0000  1.0400
136    2  2  2   0.0000  27.9213  5.8635  0.0000  0.0000   0.0000  1.0400
137    1  1  3  94.2903  59.3577  5.0000  0.0000  1.8747   1.0000  1.3538
138    3  1  3  86.8971  23.9117  5.0000  0.0000  0.0499   1.0000  2.8723
139    1  3  1  49.8324  39.3271  5.0000  0.0000  3.0156   0.0000  2.7534
140    1  3  2  70.1101  13.1217  4.4734  0.0000  0.8433   0.0000  3.0000
141    1  3  3  81.9029  32.2258  1.7397  0.0000  0.9888  68.1072  1.7777
142    1  2  3   0.0000  25.0000  3.0000  0.0000  1.0000   0.0000  1.0400
143    3  3  3  80.7324  30.4554  0.9953  0.0000  1.6310  50.0000  1.0783
144    2  3  3  75.6935  50.0000  2.0000  0.0000  1.0000   0.0000  1.1680
145    2  3  2  85.8000   9.8453  2.2720  0.0000  2.8635   0.0000  1.5800
146    3  2  3   0.0000  15.0000  2.8900  0.0000  0.0000   0.0000  2.8774
147    2  2  3   0.0000   8.5744  3.0000  0.0000  0.0000   0.0000  1.0421
148    4  4  4  66.2921  19.5195  0.9624  0.0000  0.1000   0.0000  1.2636
149    2  4  4  68.5501  19.4239  2.3592  0.0000  0.2029   0.0000  1.0000
150    2  4  2  70.7499  11.4850  4.6606  0.0000  1.5647   0.0000  1.0902
151    3  4  4  86.3294  18.3879  5.8529  0.0000  1.7361   0.0000  1.2310
152    2  4  3  73.6998  40.0000  1.8782  0.0000  4.0000   0.0000  1.1290
153    3  4  3  79.5581  34.9140  1.0801  0.0000  0.1632   0.0000  2.2206
154    4  3  4  82.3364   4.7350  1.3544  0.0000  1.4627   0.0000  1.0400
155    2  3  4  90.0000   6.6857  1.6689  0.0000  2.5771   0.0000  1.0400
156    3  3  4  92.1207  24.3937  0.5000  0.0000  1.7208   0.0000  3.0000
157    2  2  4   0.0000   0.0100  1.0000  0.0000  1.0000   0.0000  2.0000
158    4  2  4   0.0000   4.4216  0.8596  0.0000  0.9624   0.0000  1.0000
159    3  2  4   0.0000   5.0000  1.0000  0.0000  1.0000   0.0000  1.2500
160    3  5  3  60.0000   0.0000  1.0000  0.0000  1.0000   0.0000  1.0000
161    5  3  3  81.6233  30.0000  2.0000  0.0000  1.0000   0.0000  1.0000
162    5  3  5  67.5247   6.4512  4.0000  0.0000  1.0000   0.0000  2.8079
163    5  3  4  62.6634   8.4441  2.5120  0.0000  1.0000   0.0000  1.0000
164    1  5  3   0.0000   0.0100  1.0000  0.0000  1.0000   0.0000  1.0000
165    1  3  5  98.9874   6.7756  0.2680  0.0000  3.8836   0.0000  1.0000
166    5  1  3  99.6399   0.0100  3.8420  0.0000  2.0653   0.0000  1.8902
167    1  1  4  70.8533  23.2816  2.7470  0.0000  2.0166   0.0000  2.2246
168    1  4  1  69.9335  20.9406  1.8375  0.0000  0.2981   0.0000  2.1195
169    4  1  4  50.9317  18.9333  1.8833  0.0000  0.2981   0.0000  2.3646
170    1  4  4  69.3369  19.6964  2.0703  0.0000  1.0031   0.0000  1.0400
171    2  1  4  72.5949  13.8347  2.4952  0.0000  1.0000   0.0000  1.0400
172    1  4  2  72.5949  14.8347  2.4952  0.0000  1.0000   0.0000  1.0400
173    1  2  4   0.0000   2.5000  1.0000  0.0000  1.0000   0.0000  1.2500
174    1  1  6  35.0676  44.0627  0.4803  0.0000  0.1357   0.0000  0.6875
175    6  1  6  90.2886  41.7871  0.4774  0.0000  0.2704   0.0000  0.6548
176    1  6  1  82.4766  33.5783  0.1356  0.0000  0.2024   0.0000  0.9612
177    1  6  6   2.0340  25.1860  1.0950  0.0000  0.1315   0.0000  0.9638
```

```
178     5  6  5   75.9308   39.5963    1.5406    0.0000    5.0016    0.0000    1.2226
179     6  5  6   85.2677   38.4940    1.6737    0.0000    5.0138    0.0000    1.3107
180     6  3  6   86.4972   15.1409    1.1470    0.0000    0.2045    0.0000    2.4012
181     6  3  3   73.0506   30.3751    2.0217    0.0000    1.7756    0.0000    2.9301
182     6  4  6   79.1640   38.5410    1.3735    0.0000    4.9517    0.0000    2.2216
183    34       ! Nr of torsions; V1; V2; V3; V2(BO); vconj; n.u1; n.u2
184     1  1  1  1   -0.2500   11.5822    0.1899   -4.7057   -2.2047    0.0000    0.0000
185     1  1  1  2   -0.2500   31.2596    0.1709   -4.6391   -1.9002    0.0000    0.0000
186     2  1  1  2   -0.1770   30.0252    0.4340   -5.0019   -2.0697    0.0000    0.0000
187     1  1  1  3   -0.5000    5.0000   -0.5000   -9.0000   -1.0000    0.0000    0.0000
188     2  1  1  3   -0.5000    8.1082    0.1710   -8.1074   -1.0000    0.0000    0.0000
189     3  1  1  3   -1.4477   16.6853    0.6461   -4.9622   -1.0000    0.0000    0.0000
190     1  1  3  1   -0.2300   46.8253   -0.2848   -2.6326   -1.0000    0.0000    0.0000
191     1  1  3  2    1.2044   80.0000   -0.3139   -6.1481   -1.0000    0.0000    0.0000
192     2  1  3  1   -2.5000   31.0191    0.6165   -2.7733   -2.9807    0.0000    0.0000
193     2  1  3  2   -2.4875   70.8145    0.7582   -4.2274   -3.0000    0.0000    0.0000
194     1  1  3  3   -0.0002   20.1851    0.1601   -9.0000   -2.0000    0.0000    0.0000
195     2  1  3  3   -1.4383   80.0000    1.0000   -3.6877   -2.8000    0.0000    0.0000
196     3  1  3  1   -1.2244   77.8133   -0.4738   -4.7499   -3.0000    0.0000    0.0000
197     3  1  3  2   -2.5000   70.3345   -1.0000   -5.5315   -3.0000    0.0000    0.0000
198     3  1  3  3   -0.1583   20.0000    1.5000   -9.0000   -2.0000    0.0000    0.0000
199     1  3  3  1    0.0002   80.0000   -1.5000   -4.4848   -2.0000    0.0000    0.0000
200     1  3  3  2   -2.1289   12.8382    1.0000   -5.6657   -2.9759    0.0000    0.0000
201     1  3  3  3    2.5000  -25.0000    1.0000   -2.5000   -1.0000    0.0000    0.0000
202     2  3  3  3    0.8302   -4.0000   -0.7763   -2.5000   -1.0000    0.0000    0.0000
203     3  3  3  3   -2.5000   -4.0000    1.0000   -2.5000   -1.0000    0.0000    0.0000
204     0  1  1  0    0.0000    0.0000    0.0000    0.0000    0.0000    0.0000    0.0000
205     0  1  2  0    0.0000    0.0000    0.0000    0.0000    0.0000    0.0000    0.0000
206     0  2  2  0    0.0000    0.0000    0.0000    0.0000    0.0000    0.0000    0.0000
207     0  2  3  0    0.0000    0.1000    0.0200   -2.5415    0.0000    0.0000    0.0000
208     0  3  3  0    0.5511   25.4150    1.1330   -5.1903   -1.0000    0.0000    0.0000
209     2  4  4  2    0.0000    0.0000    0.0640   -2.4426    0.0000    0.0000    0.0000
210     2  4  4  4    0.0000    0.0000    0.1587   -2.4426    0.0000    0.0000    0.0000
211     0  2  4  0    0.0000    0.0000    0.1200   -2.4847    0.0000    0.0000    0.0000
212     1  1  1  6    0.0000   42.3172    3.4546  -22.6501   -1.7255    0.0000    0.0000
213     6  1  1  6    0.0000   75.5402   -0.7497   -4.0257   -1.7255    0.0000    0.0000
214     0  1  6  0    4.0000   45.8264    0.9000   -4.0000    0.0000    0.0000    0.0000
215     0  6  6  0    4.0000   45.8264    0.9000   -4.0000    0.0000    0.0000    0.0000
216     0  5  6  0    0.0336    0.6333   -0.0621   -0.0007   -0.0578    0.0000    0.0000
217     0  4  6  0    4.7334   44.7693    1.2361   -4.7157    0.0000    0.0000    0.0000
218     2       ! Nr of hydrogen bonds; Rhb; Dehb; vhb1; vhb2
219     3  2  3    2.1200   -3.5800    1.4500   19.5000
220     6  2  3    2.1200   -2.0000    1.4500   19.5000
```